\documentclass[prl,twocolumn,amsmath,groupedaddress,amssymb,floatfix,footinbib,bibnotes,longbibliography,superscriptaddress]{revtex4-2}

\usepackage{amsmath}
\usepackage{graphicx}
\usepackage{textcomp}
\usepackage{soul}
\usepackage{changepage}
\usepackage{fancyhdr}
\usepackage{amsthm,amssymb}
\usepackage{floatflt}
\usepackage{float}
\usepackage{array}
\usepackage[usenames,dvipsnames]{color}
\usepackage{comment}

\usepackage{lipsum}
\usepackage{braket}
\usepackage{mathtools}
\usepackage[version=4]{mhchem}
\usepackage{siunitx}

\usepackage[normalem]{ulem}

\usepackage{amsmath,amsfonts,amssymb}
\usepackage{mathrsfs}
\usepackage[english]{babel} 
\usepackage[colorlinks=true,citecolor=blue,linkcolor=magenta]{hyperref}
\usepackage{verbatim}
\usepackage{bbold}

\usepackage{pdfpages}

\makeatletter
\AtBeginDocument{\let\LS@rot\@undefined}
\makeatother

\begin{document}

%\title{Spin-liquids from local Hamiltonians on triangular lattice}
%\title{Chiral spin liquid in the effective spin model of the triangular lattice Hubbard model and arguments for its emergence}
\title{Four-Spin Terms and the Origin of the Chiral Spin Liquid in Mott Insulators on the Triangular Lattice}

%Chiral spin liquid on the triangular lattice with ring-exchange term
\author{Tessa Cookmeyer}
\email[]{tcookmeyer@berkeley.edu}
\affiliation{Department of Physics, University of California, Berkeley, CA, 94720, USA}
\affiliation{Materials Sciences Division, Lawrence Berkeley National Laboratory, Berkeley, California, 94720, USA}

\author{Johannes Motruk}
\affiliation{Department of Physics, University of California, Berkeley, CA, 94720, USA}
\affiliation{Materials Sciences Division, Lawrence Berkeley National Laboratory, Berkeley, California, 94720, USA}
\affiliation{Department of Theoretical Physics, University of Geneva, Quai Ernest-Ansermet 30, 1205 Geneva, Switzerland}

\author{Joel E. Moore}
\affiliation{Department of Physics, University of California, Berkeley, CA, 94720, USA}
\affiliation{Materials Sciences Division, Lawrence Berkeley National Laboratory, Berkeley, California, 94720, USA}

\begin{abstract}
At strong repulsion, the triangular-lattice Hubbard model is described by $s=1/2$ spins with nearest-neighbor antiferromagnetic Heisenberg interactions and exhibits conventional 120$^\circ$ order. Using the infinite density matrix renormalization group and exact diagonalization, we study the effect of the additional four-spin interactions naturally generated from the underlying Mott-insulator physics of electrons as the repulsion decreases.  Although these interactions have historically been connected with a gapless ground state with emergent spinon Fermi surface, we find that at physically relevant parameters, they stabilize a chiral spin-liquid (CSL) of Kalmeyer-Laughlin (KL) type, clarifying observations in recent studies of the Hubbard model. We then present a self-consistent solution based on a mean-field rewriting of the interaction to obtain a Hamiltonian with similarities to the parent Hamiltonian of the KL state, providing a physical understanding for the origin of the CSL.

\end{abstract}

\maketitle

\textit{\textbf{Introduction--}}
 %The Kalmeyer-Laughlin (KL) state \cite{Kalmeyer1987} is a chiral spin-liquid (CSL) that was introduced as a bosonic analog to the fractional quantum Hall state characterized by a fractional spin-Hall effect and topological degeneracy on the torus. Finding a parent spin Hamiltonian of the KL state \cite{nielsen2013,Thomale2009} and its generalizations, the Read-Rezayi states \cite{Glasser2015,Greiter2014}, has been of considerable interest. Generally, the parent Hamiltonians derived from conformal-field theoretic arguments have long-ranged interactions, but a short-ranged Hamiltonian can be found if only short-ranged coefficients are kept, made uniform, and tuned  \cite{nielsen2013,Glasser2015,Gong2017, nielsen2013, Bauer2014, Wietek2017}. In these cases, though, the Hamiltonian explicitly breaks time-reversal (TR) symmetry. A notable exception is seen on  the Kagome lattice near a classical chiral phase transition \cite{gong2014, Wietek2015,Messio2012}. 
 %
 %Remarkably, a recent study of the Hubbard model on the triangular lattice at half-filling found a CSL resembling the KL state between the metallic and magnetically-ordered states \cite{szasz2020}. The so-called non-magnetic insulating state has been found in numerous previous studies\cite{Sahebsara2008,Clay2008, Shirakawa2017,Laubach2015, Yang2010,Yoshioka2009,Morita2002,Kyung2006,Tocchio2009,Antipov2011, Misumi2017}, but many alternate descriptions were proposed. 
 %
 The triangular lattice has played a prominent role in the physics of spin liquids ever since they were first proposed by Anderson~\cite{Anderson1973}, and many of the candidate materials exhibit this lattice geometry~\cite{Shimizu2003,Itou2008,Li2015,Shen2016,Law2017,Ribak2017,Klanjsek2017,Zeng2020,Li2020,Sarkar2020}. In particular, some organic charge transfer salts~\cite{Shimizu2003,Itou2008} and 1$T$-TaS${}_2$~\cite{Law2017,He2018} are believed to be described by the Hubbard model on the triangular lattice in the vicinity of the Mott transition.  While the existence of a non-magnetic insulating (NMI) phase in the Hubbard model has been observed in numerous studies~\cite{Morita2002,Kyung2006,Sahebsara2008,Clay2008, Yang2010, Yoshioka2009, Tocchio2009,Antipov2011, Laubach2015, Shirakawa2017, Misumi2017}, the determination of the type of spin liquid phase in direct studies of the Hubbard model has long been elusive.
 
The problem has instead often been investigated via an effective spin model.  Deep in the insulating phase of the Hubbard model, a nearest-neighbor Heisenberg model is sufficient and contains long-ranged three-sublattice order~\cite{husetriangular,Bernu1994,Capriotti1999}.
To describe physics closer to the Mott transition, one includes a four-spin ring exchange part in addition to the Heisenberg term, a description coming from the lowest order $t/U$ expansion of the Hubbard model~\cite{MacDonald1988}. In a seminal paper, Motrunich showed using variational Monte Carlo simulations that a spin liquid with spinon Fermi surface (SFS) is a strong competitor for the ground state if the ring exchange term is large enough~\cite{Motrunich2005}. Indications for this state, in subsequent works also referred to as spin-Bose metal, have been seen in other studies including some with complementary methods~\cite{Sheng2009,Yang2010,Block2011,Mishmash2013,He2018,Zhao2021}, but remain under debate~\cite{Aghaei2020}. However, recent work on the Hubbard model suggested that the NMI is instead a chiral spin liquid (CSL) of Kalmeyer-Laughlin (KL) type~\cite{Kalmeyer1987,szasz2020,Zhu2020,Szasz2021,Chen2021}, seemingly at odds with the results for the effective spin model.

In this Letter, using a combination of exact diagonalization (ED) and infinite density matrix renormalization group (iDRMG)~\cite{Mcculloch2008} simulations, we first show that the KL spin liquid is indeed the ground state of the effective spin model around the parameter regime relevant for the Hubbard model.  We demonstrate that this CSL does not emerge as a competing state to the SFS, but rather appears at a different value of the four-spin interaction; this is to our knowledge the first demonstration of a KL ground state in a time-reversal invariant spin model on the triangular lattice.  However, we also find that much of the region which had been attributed to the SFS in previous works is occupied by a magnetically ordered zigzag state.  The second main result is to connect analytically the four-spin term, which preserves time-reversal symmetry (TRS), back to the TRS-breaking parent Hamiltonians of the KL state \cite{Thomale2009,nielsen2013} by mean-field arguments.  Hence one aspect of our work clarifies the relation between the appearance of the CSL in the triangular lattice Hubbard model and the corresponding spin model, while the second clarifies why the CSL appears in the spin model via a connection to known TRS-breaking parent Hamiltonians for the CSL.

Finding a parent spin Hamiltonian of the KL state \cite{Thomale2009,nielsen2013} and its generalizations, the Read-Rezayi states \cite{Greiter2014,Glasser2015}, has been of considerable interest. Generally, the parent Hamiltonians derived from conformal-field theoretic arguments have long-ranged interactions, but a local Hamiltonian can be found if only short-ranged coefficients are kept, made uniform, and tuned  \cite{nielsen2013,  Bauer2014, Glasser2015,Gong2017, Wietek2017,Hickey2017}. While the underlying Hamiltonian for a material in zero applied field should respect TRS, these parent Hamiltonians explicitly break TRS.  A notable exception is on the Kagome lattice near a classical chiral phase transition \cite{Messio2012,gong2014, He2014, Wietek2015,Hu2015Variational,Gong2015Global}, but no TRS-preserving spin Hamiltonian with KL ground state on the triangular lattice is known analytically.

 %It was also shown through exact-diagonalization \cite{Yang2010} that the NMI state could be studied with the $t/U$ expansion of the Hubbard model \cite{MacDonald1988}. In addition to next-nearest and next-next-nearest neighbor Heisenberg terms, The next-to-leading-order term in the expansion contains the following four-spin term.
 
\begin{figure}
    \centering
    \includegraphics[width=0.5\textwidth]{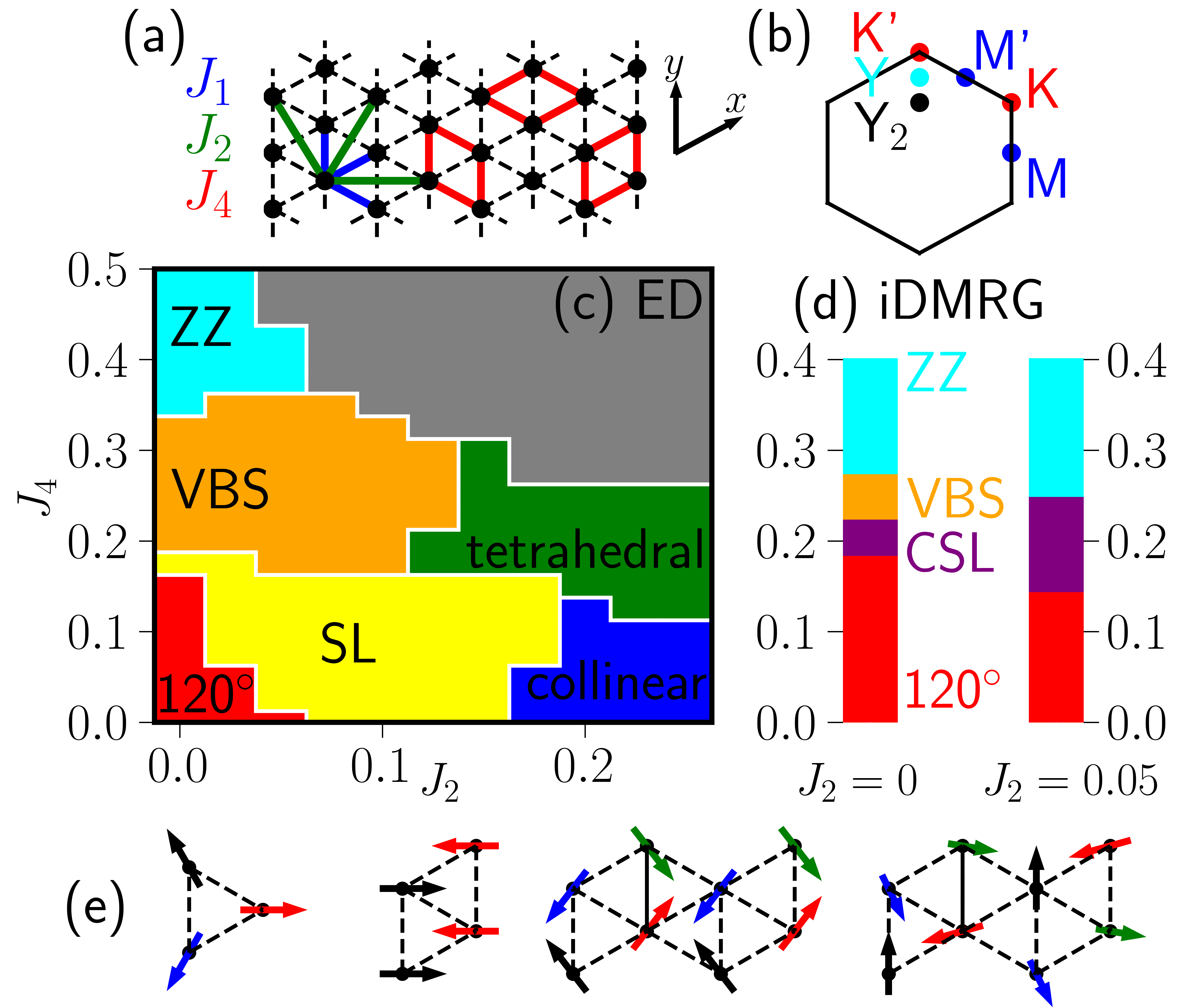}
    \caption{\textbf{(a)}  The different colored lines connect the spins involved in the different terms of Eq.~\eqref{eq:Ham} \textbf{(b)} The first Brillouin zone of the lattice showing several named points. \textbf{(c)} The proposed phase diagram from our ED results using the various orders in Fig.~\ref{fig:ED_fig_8panel}. For phase descriptions, see \cite{SM}. 
    %The 120${}^\circ$ (tetrahedral) ordering of the spins has a three-spin (four-spin) unit cell with the spins pointing towards the vertices of a regular triangle (tetrahedron). The collinear (zigzag) ordering has the spins parallel in columns (zigzags) and antiparlllel along adjactent columns (zigzags). Finally, the valence-bond solid has every other spin along one of the nearest-neighbor directions paired with its nearest neighbor. 
    The phase boundaries were determined via the symmetry sector of the ground state and first excited state \cite{SM, Wietek2017}. %The cutoffs for each phase are somewhat arbitrary and approximate. In particular, the cutoff was chosen so that there is a spin-liquid (SL) region that overlaps with the $J_1-J_2$ SL found in the literature. 
    The greyed out region is within the SFS parameter space found in \cite{Motrunich2005}, but also might have some dimer or plaquette ordering.  \textbf{(d)} The phase diagram from the iDMRG results on the $L_y=6$ cylinder on the $J_2=0.00, 0.05$ slices, which includes the CSL perhaps suggested by ED. \textbf{(e)} From left to right, the 120$^\circ$, collinear, zigzag, and tetrahedral (whose spins, connected tail-to-tail, form a tetrahedron) classical spin orders are shown \cite{SM}. }
    \label{fig:conventions}
\end{figure}

 \textit{\textbf{Model--}}
 Motivated by the $t/U$ expansion of the Hubbard model, we consider the following Hamiltonian
 \begin{equation} \label{eq:Ham}
\begin{aligned}
    H = J_1\sum_{\langle i j\rangle } \pmb S_i \cdot \pmb S_j + J_2 &\sum_{\langle \langle i j\rangle \rangle} \pmb S_i \cdot  \pmb S_j + H_4,
\end{aligned}
\end{equation}
where $\langle ij\rangle$ $(\langle \langle i j \rangle \rangle)$ denotes (next-)nearest neighbor pairs. The four-spin interaction $H_4$ is given by
  \begin{equation}
 \begin{aligned}
     H_4 = J_4 & \sum_{\langle i, j, k, l\rangle}\Big[ (\pmb S_i \cdot \pmb S_j)(\pmb S_k \cdot \pmb S_l) \\ &+ (\pmb S_i \cdot \pmb S_l)(\pmb S_j \cdot \pmb S_k)
    - (\pmb S_i \cdot \pmb S_k)(\pmb S_j \cdot \pmb S_l) \Big].
    \label{eq:HamJ4}
\end{aligned}
 \end{equation} where $\langle i,j,k,l\rangle$ denotes a sum over unique rhombuses as defined by unique next-nearest neighbor pairs $\langle \langle i k\rangle \rangle$ (see Fig.~\ref{fig:conventions}). 
 This four-spin term is related to the extensively studied ring-exchange operator \cite{misguich1998,kubo1998,Misguich1999,LiMing2000,Motrunich2005,Fuseya2009,Sheng2009,Grover2010,Block2011,Mishmash2013,Holt2014,He2018,riedl2019,Seki2020, Aghaei2020} via the $4J_2 = J_4$ line. Furthermore, studies on the $J_4=0$ line have focused on the emergence of a ``$J_1$-$J_2$ spin liquid''~\cite{Kaneko2014,Zhu2015,Hu2015,Gong2017,Wietek2017,Saadatmand2017,Gong2019,Hu2019}. Treated classically, the Hamiltonian exhibits spontaneous TRS breaking into a tetrahedrally ordered phase \cite{Korshunov1993,kubo1997ground,Momoi1997,Messio2011, SM} %\cite{Wang2006,Pollmann2009,fehske2007,Wietek2017a, white2005,Crosswhite2008,Motruk2016,Cincio2013,Tu2013,Zaletel2013,Zaletel2015}
 further motivating this particular model. From here on in, we take $J_1=1$ and $\sum_i S_i^z=0$.

 \textit{\textbf{Exact diagonalization--}}
 We perform ED on $6 \times 4$ spins with periodic boundary conditions (PBC). The PBC are chosen such that the unit cell is translated in the $\hat y$ direction and in the $2\hat x-\hat y$ direction. We compute the structure factor for the spin, $\pmb S_i$, and dimer, $D_{\pmb \alpha}^{\pmb x_i} =\pmb S_{\pmb x_i} \cdot \pmb S_{\pmb x_i+\pmb \alpha}$, correlations
 \begin{equation}
     S(\pmb q) =  \sum_{i,j} \left(\langle \pmb S_{i} \cdot \pmb S_{j} \rangle - \langle \pmb S_{i}\rangle \cdot \langle \pmb S_{j}\rangle  \right)e^{i\pmb q\cdot( \pmb x_j - \pmb x_i)}
 \end{equation}
  \begin{equation}
  \begin{aligned}
     D_{\pmb \alpha}(\pmb q) &= \sum_{i,j} \big( \langle  D_{\pmb  \alpha}^{\pmb x_i}D_{ \pmb \alpha}^{\pmb x_j}\rangle- \langle D_{\pmb  \alpha}^{\pmb x_i} \rangle \langle D_{ \pmb \alpha}^{\pmb x_j} \rangle \big) e^{i\pmb q\cdot( \pmb x_j - \pmb x_i)}
\end{aligned}
 \end{equation}
with $\pmb \alpha$ being the vector to one of the three nearest neighbors, and $\pmb S_{\pmb x_i}$ is an alternative notation for $\pmb S_i$. Large values of $S(\pmb q)$ and/or $D_{\pmb \alpha}(\pmb q)$ indicate ordered phase;  see \cite{SM} for more information   about the various orders.

%Because the system is not square, the order is not rotationally symmetric.
%, and, for instance, the collinear order has a peak at $S(M)$ but not at the other rotationally symmetric points in the Brillouin zone. 
To distinguish the tetrahedral from the collinear state, we compute a nematic order parameter, a chiral-chiral order parameter, \cite{Wietek2017} and we study the effect of adding a small TRS-breaking term to the Hamiltonian. As shown in the Supplemental Material~\cite{SM}, this analysis clearly shows that large $S(M')$ [$S(M)$] is indicative of  tetrahedral [collinear] order.
%To distinguish tetrahedral from the collinear state, we  introduce a small $H_\chi = J_\chi\sum_{\triangle,\triangledown} \pmb S_i\cdot (\pmb S_j \times \pmb S_k)$ term to the Hamiltonian and compute $\Delta \chi/J_\chi$ where $\chi = \langle \pmb S_i \cdot (\pmb S_{j}\times \pmb S_k)\rangle$ with $i, j, k$ going clockwise around a triangle (and $\langle \cdot \rangle$ denotes the expectation averaged over all triangles in the lattice). Since the tetrahedral state breaks TRS, it would have a much larger response \cite{Gong2017,Wietek2017}. As shown in the SM, this analysis clearly shows that the order at $S(M')$ is indicative of the tetrahedral order and $S(M)$ is indicative of collinear order.

\begin{figure}
    \centering
    \includegraphics[width=0.5\textwidth]{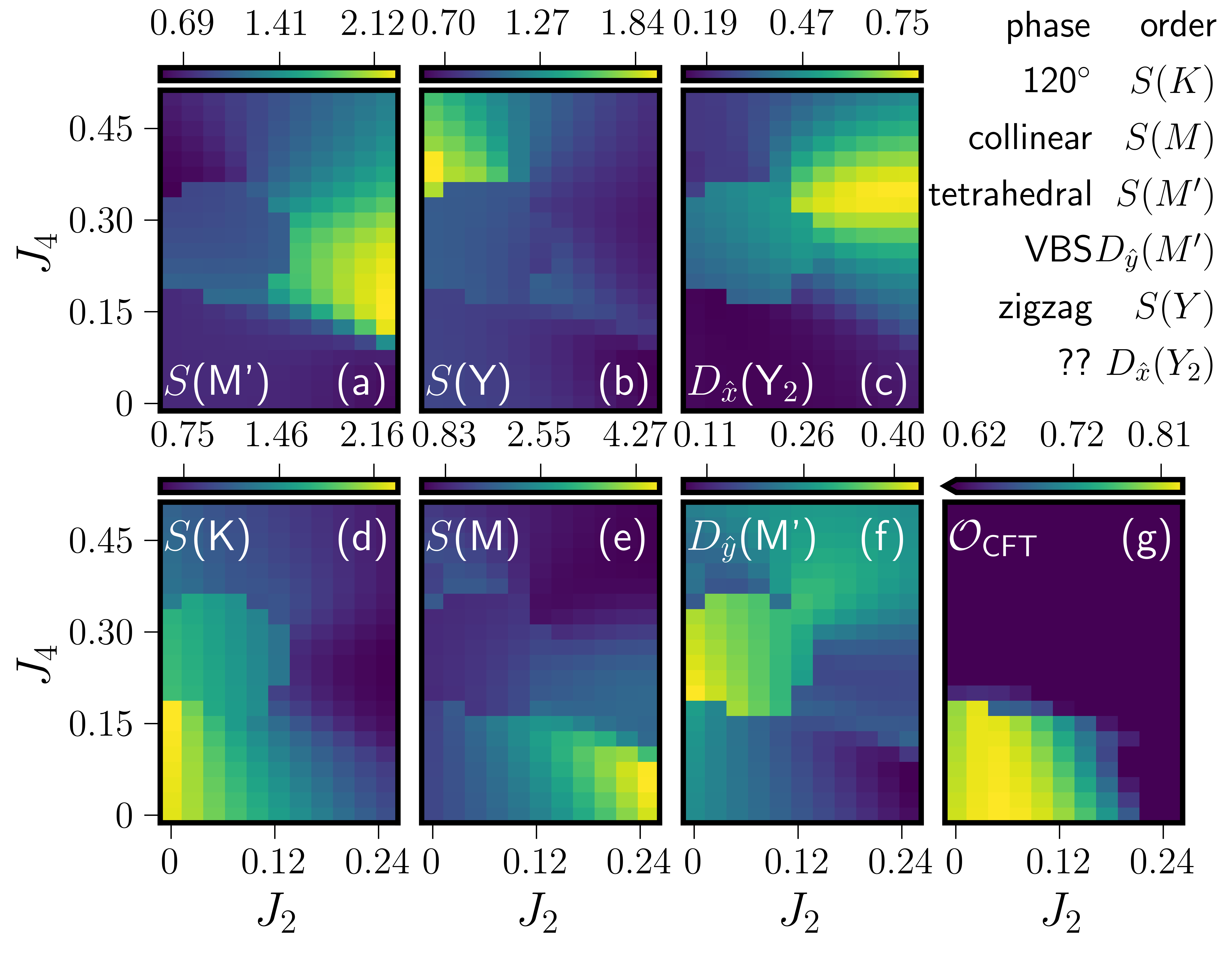}
    \caption{ \textbf{(a)-(f)} Various orders are shown in color vs. $J_2$ and $J_4$. The table in the upper right indicates the phase that each order corresponds to. \textbf{(g)} The overlap of the ground state with the manifold of KL states, which suggests that the CSL may appear for small $J_2$ and $J_4$.}
    \label{fig:ED_fig_8panel}
\end{figure}

Additionally, we are most interested in checking whether the chiral spin-liquid phase appears. For that reason, we compute $\mathcal O_{\text{CFT}} = \sqrt{\sum_{i=1}^4 |\langle \psi | \text{KL}_i\rangle |^2}$, the overlap of the ground state with its projection into the subspace spanned by the four orthonormalized KL states, $|\text{KL}_i\rangle$ (given explicitly in Ref.~\cite{Nielsen2014}). The degeneracy comes from a combination of twofold topological and TRS breaking degeneracy each.

\begin{figure}
    \centering
    \includegraphics[width=0.5\textwidth]{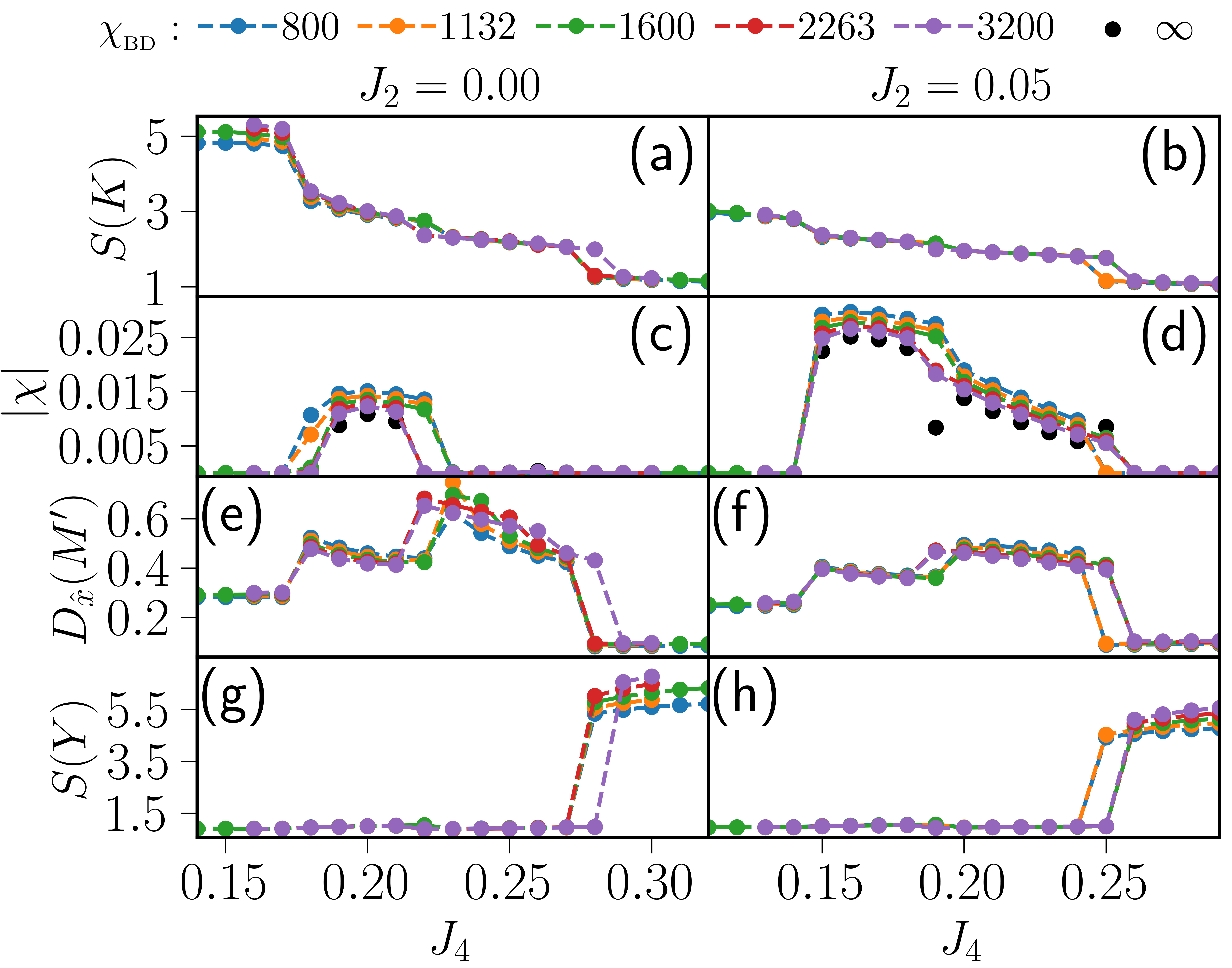}
    \caption{We plot various order parameters that we extract from ground state wave-function from iDMRG for the $L_y=6$ cylinder, and $J_2=0$ ($J_2=0.05$) for the left (right) column respectively.  In \textbf{(a)-(b)} [\textbf{(g)-(h)}] We plot the spin-spin correlation at the $K$ [$Y$] point respectively. We see a jump in the value corresponds to a phase boundary. \textbf{(c)-(d)} We plot $\chi= \langle \pmb S_i \cdot (\pmb S_j \times \pmb S_k)\rangle$ averaged over all triangles of the lattice from the iDMRG results at varying bond dimension, $\chi_{{}_{\textrm{BD}}}$. In \textbf{(d)}, the jump in the non-zero value of $\chi$ at  $J_4=0.19$ corresponds to whether the trivial ($J_4\le 0.19$) or semion ($J_4\ge0.20$) sector of the KL state is the ground state as evidenced by the entanglement spectra. We include an extrapolation \cite{SM} to $\chi_{{}_{\textrm{BD}}} \to \infty$ where it is non-zero. In $\textbf{(e)-(f)}$, we plot the dimer-dimer correlation at the $M'$ point for dimers in the $\hat x$ direction, which signals the VBS state. The phase boundaries estimated from this data are plotted in Fig.~\ref{fig:conventions}.}
    \label{fig:IDMRG}
\end{figure}

From all of the data presented in Fig.~\ref{fig:ED_fig_8panel}, we see that there are potentially many ordered states, and we present a phase diagram in Fig.~\ref{fig:conventions}(c). 
%It is worth noting that 
%Ref.~\cite{Mishmash2013} finds a spin-liquid for $J_2=0$ and $J_4\gtrsim 0.3$, but we observe that the zigzag state appears as the ground state in ED. We will also address this question in iDMRG below. 
Most interesting, however, is that in the region most relevant for the Hubbard model at small $J_2$ and $J_4 \sim 0.1-0.15$, the overlap with the CSL is large. 

\textit{\textbf{iDMRG--}}
In order to investigate this tendency on larger system sizes, we focus on the region with $J_2 \le 0.05$ and $J_4 \le 0.4$ and study it with iDMRG. We consider the model on infinite cylinders of circumferences $L_y = 6$ and $8$ sites and compute the ground state
%The density matrix renormalization group (DMRG) method \cite{White1992} is a variational method on a class of states called matrix product states (MPS) \cite{Schollwock2011}. In the infinite DMRG (iDMRG) method \cite{??}, the MPS is assumed to be translationally invariant resulting in an algorithm for, in-principle, an infinite system whose unit cell is described by the MPS. For a two-dimensional system, the unit cell is taken to be $L_x$ rings on a cylinder of height $L_y$. In this work, we always take $L_x=2$ unless explicitly noted. iDMRG has a tuning parameter, the bond dimension $\chi_{{}_{\textrm{BD}}}$, that limits the size of the MPS; when $\chi_{bd}\to \infty$, the method is exact.
%
on the slices $J_2=0$ and $J_2=0.05$ at various bond dimension $\chi_{{}_{\textrm{BD}}}$. 
% For these slices, we use a unit cell that is two rings wide with cylinder height $L_y=6$ or $L_y=8$. 
We use the \texttt{TeNPy} library \cite{tenpy} and give further details of the numerics in the Supplemental Material~\cite{SM}. 
%To perform the iDMRG at a point $(J_2,J_4)$, we create an initial state by starting with a random product state and then do two sweeps with an added small time-reversal (TR) breaking term and a maximum bond dimension of 20. We then set the maximum bond dimension to $\chi_\text{{}_{\textrm{BD}}}$, and we perform 40 sweeps with a mixer, and then perform sweeps until the energy of the state has converged to $10^{-8}$. 
%
The results for the $L_y=6$ cylinder are presented in Fig.~\ref{fig:IDMRG} and are summarized in Fig.~\ref{fig:conventions}(d). We find similar phases as in ED. The spins order into the 120${}^\circ$ (zigzag) state at low (high) $J_4$, respectively. At intermediate $J_4$, we find a phase that breaks TRS by acquiring a non-zero value of the chiral order parameter %$\chi$
$\chi = \langle \pmb S_i \cdot (\pmb S_{j}\times \pmb S_k)\rangle$ with $i, j, k$ going clockwise around a triangle (and $\langle \cdot \rangle$ denotes the expectation averaged over all triangles in the lattice),
which we identify as the KL CSL below.  Furthermore, we confirm the presence of the valence-bond solid (VBS) on the $J_2=0$ slice reported in Ref.~\cite{He2018}.

For the $L_y=8$ cylinder, we focus on demonstrating that, at the point $(J_2,J_4)=(0.05,0.18)$, the ground state is the CSL. By running the algorithm at different $(J_2,J_4)$, we find the same states as in the $L_y=6$ cylinder. In addition to an unbiased run, we use those states as the initial state to bias the algorithm towards converging to a non-CSL state at $(0.05,0.18)$. By $\chi_{{}_{\textrm{BD}}}=1600$, however, the algorithm always converges to the CSL, and an unbiased run with $\chi_{{}_{\textrm{BD}}}=3200$ also finds the CSL.

\textit{Identification as the CSL--} %As mentioned above, we find a phase which spontaneously breaks TRS. 
Here, we identify the TRS breaking phase as the Kalmeyer-Laughlin state by studying the entanglement spectrum and performing a spin-Hall numerical experiment.
\begin{figure}
    \centering
    \includegraphics[width=0.5\textwidth]{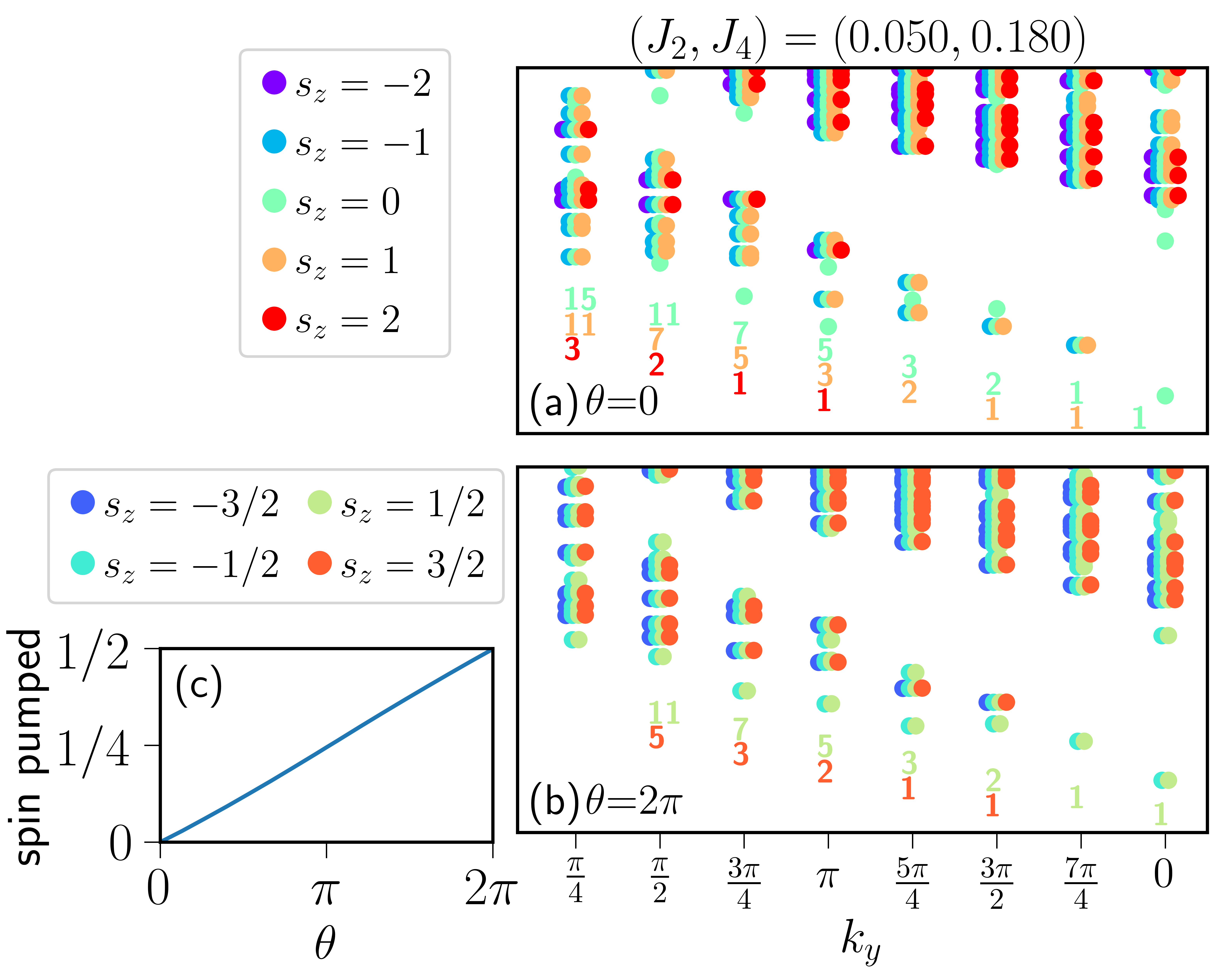}
    \caption{\textbf{(a)} We plot the entanglement spectrum for the ground state at $(J_2,J_4)=(0.05,0.18)$ on the $L_y=8$ cylinder with $\chi_{{}_{\textrm{BD}}}=1600$. The $y$-axis is $-s\ln(s)$ where $s$ are the Schmidt values. The color indicates the charge as specified in the legend, and different charges are offset slightly from each other to more clearly show the degeneracy. For each momentum, the counting of the lowest cluster of  Schmidt values is shown for each of the $s_z\ge 0$ charges in color. They show the correct pattern for the Kalmeyer-Laughlin state. \textbf{(b)} We make the same plot as in (a) after adiabatically inserting one flux quantum through the cylinder. Although the Hamiltonian is the same, the entanglement spectrum has changed indicating a topological degeneracy of the state. \textbf{(c)} During the flux insertion, we can monitor how much spin has flowed along the cylinder. We see that exactly a spin-1/2 is pumped across the system indicating a quantized fractional spin-Hall effect.}
    \label{fig:spinpump_fig}
\end{figure}
We focus on $(J_2,J_4)=(0.05,0.18)$, and show the results of both in Fig.~\ref{fig:spinpump_fig}. First, we compute the entanglement spectrum, which shows the correct counting  for the KL state; each of the levels with spin quantum number $|s_z|\in \{0,1,2\}$ show the degeneracy pattern of $1,1,2,3,5,\ldots$ as we move around the momentum~\cite{Wen1991,Li2008}. Next, we thread flux through the cylinder by replacing $S_i^+S_j^- \to S_i^+S_j^-e^{i\theta (y_i-y_j)/L_y}$, so that, upon going around the cylinder, a spin will have picked up a phase of $e^{i\theta}$. 
% As we change $\theta$, we adiabatically update the ground state and the environment. 
As can be seen in Fig.~\ref{fig:spinpump_fig}(c), adding $2\pi$ flux moves exactly $1/2$ a spin along the cylinder. Additionally, although the Hamiltonian has returned to the original Hamiltonian up to a gauge transformation, the ground state has a different entanglement spectrum with half-integer spin quantum numbers. Indeed, inserting $2\pi$ flux exchanges the trivial and semion sectors of the ground state manifold \cite{footnote1} of the KL state on the infinite cylinder, which is precisely what we see in this numerical experiment. 

\textit{Zigzag vs. spinon Fermi surface--}
In a recent DMRG study of Eq.~\eqref{eq:Ham} at $J_2=0$, the authors of Ref.~\cite{He2018} find a spin-liquid at $J_4\gtrsim 0.3$ that they identify as a spinon Fermi surface (SFS) phase. We instead find that a zigzag ordered state at finite bond dimension has lower energy for the parameter choices we studied (i.e. $J_4 \le 0.4$) consistent with our ED results. By biasing the initial state towards the SFS or zigzag state, we compare how the energy depends on the truncation error of DMRG at the point $J_4=0.4$ ~\cite{Legeza1996,Hubig2018,SM} which allows us to estimate the ground state energy at infinite bond dimension. However, we still find the zigzag state is preferred for the $L_y=6$ cylinder where we performed the analysis. Future work may attempt to clarify whether the SFS appears at other points in the parameter space; a recent effort in that direction is seen in \cite{Aghaei2020}. Regardless, the SFS does not seem to be favored in the regime most physically close to the Hubbard model. These results could also be investigated by variational Monte Carlo studies since previous works seem not to have considered a trial state with zigzag order~\cite{Motrunich2005,Grover2010,Mishmash2013}.

\textit{\textbf{Discussion}} As mentioned in the introduction, this spin-model is motivated by the Hubbard model's $t/U$ expansion. In particular, at order $t^4/U^3$, the Hubbard model gives $J_1 = 4(1-7t^2/U^2) t^2/U$, $J_2 =4t^4/U^3$, $J_3 = 4 t^4/U^3$, and $J_4 = 80t^4/U^3$ where $J_3$ is a next-next-nearest-neighbor Heisenberg interaction \cite{MacDonald1988}. Ignoring $J_3$, if we use the value of $ U/t\sim10.6$ for the transition to the CSL phase from Ref.~\cite{szasz2020}, we would estimate the transition to be at $(J_2,J_4) \sim (0.01,0.19)$, essentially where we find it.

One could still ask why the KL state should be the ground state for the Hamiltonian Eq.~\eqref{eq:Ham}, though. In this section, we connect the above Hamiltonian to the parent Hamiltonians of Refs.~\cite{nielsen2013,Glasser2015,Greiter2014}.
In the Supplemental Material \cite{SM}, we derive that, for spin-1/2s, we can rewrite Eq.~\eqref{eq:HamJ4} as
\begin{equation}
\begin{aligned}
H_4 &=-\frac{107}{88}J_4 \sum_{\langle i j \rangle} \pmb S_i \cdot \pmb S_j + 3NJ_4\frac{129}{352} \\
+ &J_4 \sum_{\langle i,j,k,l\rangle}\left[-\frac{39}{88} \hat \chi_{ijkl}^2 - \frac{21}{22} (\hat \chi_{ijkl}^2)^2 + \frac{8}{11} (\hat \chi_{ijkl}^2)^3 \right],
\end{aligned}
\label{eq:Ham_preMFT}
\end{equation}
where $\hat \chi_{ijkl}^2 = \mathcal O_\triangle(i,j,l) \mathcal O_\triangledown(k,l,j) +\mathcal O_\triangledown(k,l,j)\cdot \mathcal O_\triangle(i,j,l)   $ for $\mathcal \mathcal{O}_{\triangle/\triangledown}(i,j,k)= 2 \pmb S_i \cdot (\pmb S_j \times \pmb S_k)$ and $N$ is the number of sites. 

We now mean-field decouple $(\hat \chi_{ijkl}^2)^n$. In the phase we are looking for, the scalar chirality $\chi = \langle O_\triangle(i,j,k)\rangle/2=\langle O_\triangledown(i,j,k)\rangle/2$ takes a non-zero value on all triangles. Rewriting $ \mathcal O_{\triangle/\triangledown}/2= \chi + \epsilon_{\triangle/\triangledown}$, expanding, and keeping only to order $\epsilon$, we arrive at the Hamiltonian
\begin{equation}
\begin{aligned}
H 
=\left(J_1-\frac{107}{88}J_4\right) \sum_{\langle i j \rangle} \pmb S_i \cdot \pmb S_j + J_2 \sum_{\langle \langle i j \rangle \rangle} \pmb S_i \cdot \pmb S_j &\\
+ 3NJ_4 \frac{129}{352}+3N J_4 \left[ \frac{39}{11}\chi^2 +\frac{63}{22}8^2 \chi^4 - \frac{5}{11}8^4\chi^6\right]& \\
+\underbrace{3J_4\left[-\frac{39}{11}\chi - \frac{21}{11}8^2 \chi^3 + \frac{3}{11} 8^4 \chi^5\right]}_{J_\chi} \sum_{\triangle,\triangledown}\pmb S_i \cdot (\pmb S_j \times \pmb S_k).&
\label{eq:MFT}
\end{aligned}
\end{equation}
 By adjusting $J_4$ and $J_2$, we are essentially following the program of localizing the long-range parent Hamiltonian of Refs.~\cite{nielsen2013,Thomale2009,Glasser2015,Greiter2014}; however, we also have self-consistency conditions. In semi-quantitative agreement with the iDMRG results (Fig.~\ref{fig:conventions}(d)), we show that when $J_2/[J_1-(107/88)J_4]=0.05$ the point  $J_4=0.13$ produces a self-consistent solution  with $\chi \approx -0.116$ and $J_\chi/[J_1-(107/88)J_4] \approx 0.268$ \cite{SM}, whose ground state is known to be the KL state \cite{Wietek2017,Gong2017}. We note that the mean-field decoupling happens only on the level of the chiral order parameter and the ground state of the resulting Hamiltonian \eqref{eq:MFT} still has to be found by iDMRG.
 
 Further evidence in support of the validity of this rewriting comes from the similarity of the phase diagram of Eq.~\eqref{eq:Ham} at intermediate $J_4$ in comparison to the phase diagram of the $J_1$-$J_2$-$J_\chi$ Hamiltonian at intermediate $J_\chi$ studied in Refs.~\cite{Wietek2017,Gong2017}. In particular, we find the three most relevant competing phases for $J_4=0.16$ are the 120${}^\circ$ order, the CSL, and the tetrahedral order \cite{SM}, in analogy to $J_\chi \sim 0.2$.
 Additionally, the rewriting in Eq.~\eqref{eq:Ham_preMFT} is reminiscent of the analysis in Ref.~\cite{Baskaran1989} where the nearest neighbor term is rewritten as related to $[\pmb S_i \cdot (\pmb S_j \times \pmb S_k)]^2$. The author then writes down and analyzes a free-energy expression to argue that TRS is spontaneously broken when $J_2 \ne 0$. Although that is not seen in numerics, future work could apply a similar analysis to our Eq.~\eqref{eq:Ham_preMFT}. 
 
 \textit{\textbf{Conclusion--}} We have demonstrated that a CSL appears in the effective spin model for the Hubbard model on the triangular lattice at half-filling in the parameter space near the physically relevant region. Furthermore, through a rewriting of Eq.~\eqref{eq:Ham}, we heuristically argued that the CSL emerges in this model because the four-spin term favors spontaneous TRS breaking, after which the mean-field Hamiltonian resembles known parent Hamiltonians of the KL state. This result provides some understanding of the origin of the CSL in the Hubbard model found in Refs.~\cite{szasz2020,Chen2021}.  We additionally have found that the SFS may only be the ground state in a more restricted part of the phase diagram than previously thought.  Beyond the triangular lattice, the approach of seeking self-consistent numerical solutions of a mean-field-decoupled Hamiltonian could potentially aid in understanding the appearance of spin liquids in some other situations.
 
 \textit{Note added--}A recent preprint \cite{zhang2021}, using a heuristic Schwinger boson argument, may provide an alternative understanding of the origin of the KL state in this model.
 
 \textit{Acknowledgements--} We thank Aaron Szasz and Mike Zaletel for helpful conversations and collaboration on related work. This work was supported as part of the Center for Novel Pathways to Quantum Coherence in Materials, an Energy Frontier Research Center funded by the U.S. Department of Energy, Office of Science, Basic Energy Sciences.  T.C. was supported by NSF DGE 1752814 and NSF DMR-1918065. J.M. received funding through DFG research fellowship No.\ MO 3278/1-1 and TIMES at Lawrence Berkeley National Laboratory, supported by the U.S. Department of Energy, Office of Basic Energy Sciences, Division of Materials Sciences and Engineering, under Contract No.\ DE-AC02-76SF00515. J.E.M. acknowledges support from a Simons Investigatorship.  Numerical computations were performed on the Lawrencium cluster at Lawrence Berkeley National Laboratory.


%\bibliography{pf.bib}
% 
% 
\bigskip

\onecolumngrid
\newpage



\section*{Supplementary material}

# Supplemental Material for "Four-Spin Terms and the Origin of the Chiral Spin Liquid in Mott Insulators on the Triangular Lattice"

Tessa Cookmeyer,[1, 2, *] Johannes Motruk,[1, 2, 3] and Joel E. Moore[1, 2]

[1]*Department of Physics, University of California, Berkeley, CA, 94720, USA*
[2]*Materials Sciences Division, Lawrence Berkeley National Laboratory, Berkeley, California, 94720, USA*
[3]*Department of Theoretical Physics, University of Geneva,*
*Quai Ernest-Ansermet 30, 1205 Geneva, Switzerland*

In this Supplemental Material, we include a brief section on the different notation used in the literature and how our model is related to the Hubbard model. We then produce a classical phase diagram, and discuss the various orders that appear both classically and quantum mechanically. We discuss some details of exact-diagonalization (ED) and infinite density matrix renormalization group (iDMRG) as well as show some data not shown in the main text. Finally, we elaborate on the mean-field argument presented in the main text.

## NOTATION AND CONNECTION TO THE HUBBARD MODEL

We specify the Hamiltonian via $J_1$, $J_2$, and $J_4$, as in

$$H = J_1 \sum_{\langle i,j \rangle} \boldsymbol{S}_i \cdot \boldsymbol{S}_j + J_2 \sum_{\langle\langle i,j \rangle\rangle} \boldsymbol{S}_i \cdot \boldsymbol{S}_j + J_4 \sum_{[i,j,k,l]} \left[ (\boldsymbol{S}_i \cdot \boldsymbol{S}_j)(\boldsymbol{S}_k \cdot \boldsymbol{S}_l) + (\boldsymbol{S}_i \cdot \boldsymbol{S}_l)(\boldsymbol{S}_j \cdot \boldsymbol{S}_k) - (\boldsymbol{S}_i \cdot \boldsymbol{S}_k)(\boldsymbol{S}_j \cdot \boldsymbol{S}_l) \right]. \quad (1)$$

with $J_1 = 1$. However, many references (e.g. [1, 2]) instead express the Hamiltonian in terms of the permutation operator $P_4$, which permutes the spins around a rhombus. The two are made equivalent by noting that for spin-1/2s [3, 4].

$$P_4 + \text{H.c.} = 4 \left[ (\boldsymbol{S}_i \cdot \boldsymbol{S}_j)(\boldsymbol{S}_k \cdot \boldsymbol{S}_l) + (\boldsymbol{S}_i \cdot \boldsymbol{S}_l)(\boldsymbol{S}_j \cdot \boldsymbol{S}_k) - (\boldsymbol{S}_i \cdot \boldsymbol{S}_k)(\boldsymbol{S}_j \cdot \boldsymbol{S}_l) \right] + \frac{1}{2}(\boldsymbol{S}_i + \boldsymbol{S}_j + \boldsymbol{S}_k + \boldsymbol{S}_l)^2 - \frac{5}{4}, \quad (2)$$

which implies

$$\begin{aligned} K_4 \sum_{[i,j,k,l]} P_4 + \text{H.c.} = &5K_4 \sum_{\langle ij \rangle} \boldsymbol{S}_i \cdot \boldsymbol{S}_j + K_4 \sum_{\langle\langle ij \rangle\rangle} \boldsymbol{S}_i \cdot \boldsymbol{S}_j + \text{ constant} \\ &+ 4K_4 \sum_{[i,j,k,l]} \left[ (\boldsymbol{S}_i \cdot \boldsymbol{S}_j)(\boldsymbol{S}_k \cdot \boldsymbol{S}_l) + (\boldsymbol{S}_i \cdot \boldsymbol{S}_l)(\boldsymbol{S}_j \cdot \boldsymbol{S}_k) - (\boldsymbol{S}_i \cdot \boldsymbol{S}_k)(\boldsymbol{S}_j \cdot \boldsymbol{S}_l) \right] \end{aligned} \quad (3)$$

With this notation change, we remake the phase diagram of [2] in Fig. S1 to allow for a more direct comparison.

Additionally, as mentioned in the main text, the Hubbard model has [5]

$$J_1 = 4(1 - 7t^2/U^2)t^2/U; \qquad J_2 = 4t^4/U^3; \qquad J_3 = 4t^4/U^3; \qquad J_4 = 80t^4/U^3 \quad (4)$$

where $J_3$ indicates the strength of a next-next-nearest-neighbor interaction that we are not taking into account. Studying the Hubbard model, then, roughly corresponds to the line $J_4 = 20J_2$. If we take the value $U/t = 10.6$ from Ref. [6] where the transition to the CSL occurs, that corresponds to a value of $(J_2/J_1, J_4/J_1) \approx (0.01, 0.19)$, which is near where we find the CSL.

## CLASSICAL PHASE DIAGRAM

This system was studied classically in Ref. [4] using the permutation notation, so the authors only studied the line $J_4 S^2 = J_2$. Considering only three and four spin unit cells, they find a transition from the 120° order to tetrahedral order at $J_4 S^2 = J_2 = 0.075$. Additionally, [7] studies the phase diagram at $J_4 \approx 0$. When $J_4 = 0$, there is a transition

---

* tcookmeyer@berkeley.edu

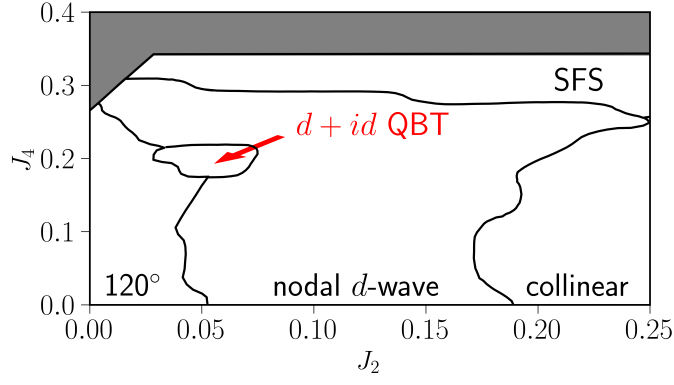

Figure S1. The phase diagram using the boundaries determined by variational Monte Carlo from Ref. [2]. See [2] for the description of the various phases. The greyed-out region was not studied, and we only plot the portion with $J_2, J_4 > 0$. The region where we find the CSL is close to the $d + id$ QBT phase.

from the 120° order to a four-spin order phase at $J_2 = 1/8$ in which any configuration with no magnetization has minimal energy. A small $J_4 > 0$ results in the tetrahedral state being preferred, which spontaneously breaks chiral symmetry.

If we open up to the entire phase diagram, the expressions for the energy of three and four spin units cells is

$$
\begin{aligned}
E_3 =& \frac{1}{2}(J_1 - J_4 S^2)\left[(\boldsymbol{S_1} + \boldsymbol{S_2} + \boldsymbol{S_3})^2 - 3S^2\right] + 3J_2 S^2 \\
&+ 2J_4\left[(\boldsymbol{S_1}\cdot\boldsymbol{S_3})(\boldsymbol{S_1}\cdot\boldsymbol{S_2}) + (\boldsymbol{S_2}\cdot\boldsymbol{S_3})(\boldsymbol{S_1}\cdot\boldsymbol{S_2}) + (\boldsymbol{S_1}\cdot\boldsymbol{S_3})(\boldsymbol{S_3}\cdot\boldsymbol{S_2})\right] \\
E_{4\triangle} =& \frac{1}{4}(J_1 + J_2)\left[(\boldsymbol{S_1} + \boldsymbol{S_2} + \boldsymbol{S_3} + \boldsymbol{S_4})^2 - 4S^2\right] \\
&+ J_4\left[(\boldsymbol{S_1}\cdot\boldsymbol{S_2})(\boldsymbol{S_3}\cdot\boldsymbol{S_4}) + (\boldsymbol{S_1}\cdot\boldsymbol{S_4})(\boldsymbol{S_3}\cdot\boldsymbol{S_2}) + (\boldsymbol{S_1}\cdot\boldsymbol{S_3})(\boldsymbol{S_2}\cdot\boldsymbol{S_4})\right] \\
E_{4\square} =& \frac{J_1 + J_2}{4}(\boldsymbol{S_1} + \boldsymbol{S_2} + \boldsymbol{S_3} + \boldsymbol{S_4})^2 - S^2 J_1 - \frac{J_2 + S^2 J_4}{2}(\boldsymbol{S_1}\cdot\boldsymbol{S_3} + \boldsymbol{S_2}\cdot\boldsymbol{S_4}) + \frac{J_4}{2}\Big[4(\boldsymbol{S_1}\cdot\boldsymbol{S_3})(\boldsymbol{S_2}\cdot\boldsymbol{S_4}) \\
&+ (\boldsymbol{S_1}\cdot\boldsymbol{S_2})(\boldsymbol{S_1}\cdot\boldsymbol{S_4}) + (\boldsymbol{S_1}\cdot\boldsymbol{S_2})(\boldsymbol{S_2}\cdot\boldsymbol{S_3}) + (\boldsymbol{S_3}\cdot\boldsymbol{S_2})(\boldsymbol{S_4}\cdot\boldsymbol{S_3}) + (\boldsymbol{S_4}\cdot\boldsymbol{S_1})(\boldsymbol{S_4}\cdot\boldsymbol{S_3})\Big]
\end{aligned}
\tag{5}
$$

where $\triangle$ and $\square$ denote the two inequivalent ways of translating the 4-spin unit cell (see Fig. S3). We then find the lowest energy configuration subjected to the constraint that the net magnetization is zero (in order to make connection with the quantum mechanical calculation).

We show the results of this minimization in Fig. S2. Without the constraint, the system will ferromagnetically order at larger $J_4$; however, with the constraint, we find that all of the spin-ordered phases we see in iDMRG appear classically. Namely, we find the 120°, collinear, tetrahedral, and zigzag phases. Note that for the collinear phase, which appears at $J_4 = 0$, any spin arrangement with zero magnetization is of minimal energy, so we label it "arbitrary", but quantum fluctuations prefer a collinear alignment [8]. We show the spin-arrangements in Fig. S3. Due to the four-spin interaction, the resulting phase-diagram is dependent on $J_2/J_1$ and $J_4 S^2/J_1$. By choosing $S^2 = 3/4$ as is the case for spin-1/2s, the phase boundaries are not that far from the quantum case.

## ORDERS

We include the schematics of the various ordered phases we describe in the main text. We include the correlation functions for the phases found in ED/iDMRG in Figs S4-S9. The units of $S(q)$ and $D_{\hat{x}}(q)$ are arbitrary but consistent between the plots.

**120°**− In this phase, there are peaks of $S(q)$ at the $K$ points in the Brillouin zone as seen in Fig. S4. This corresponds to the classical three-spin configuration seen in Fig. S3 where the spins point to the vertices of a regular triangle.

**Tetrahedral**− In this phase, there are peaks of $S(q)$ at the $M$ points in the Brillouin zone. This corresponds to the classical four-spin configuration seen in Fig. S3 where the spins point to the vertices of a regular tetrahedron. In this

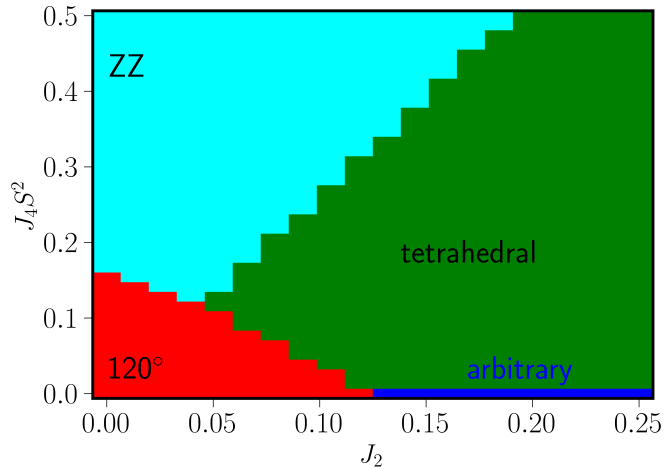

Figure S2. The classical phase diagram minimizing Eq. (5) with the constraint that the net magnetization is zero. In the arbitrary phase, any spin configuration with a four-spin unit cell, triangular tiling, and no magnetization has the same energy, but spin-1/2s prefer a collinear alignment [8]. The orderings are shown in Fig. S3.

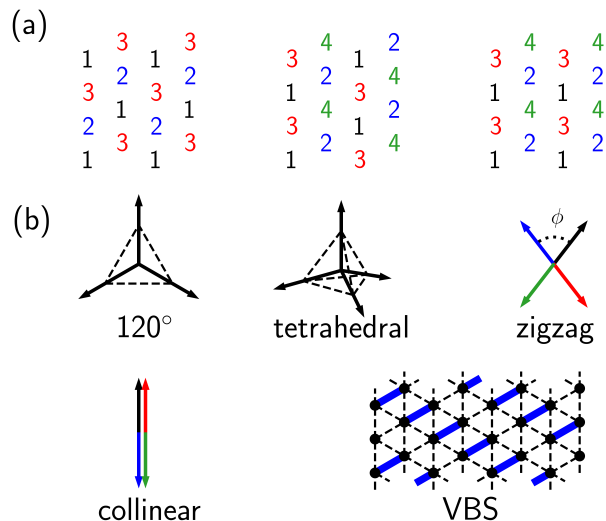

Figure S3. **(a)** How the unit cells are labeled in Eq. (5) for, reading left to right, a three-spin unit cell, four-spin unit cell with triangular tiling (4△), and four-spin unit cell with rectangular tiling (4□), respectively. **(b)** Typical ordering seen classically and in iDMRG. The arrangement of the spins correspond to the classical orderings seen. The zigzag state requires □ tiling and the color indicates which sublattice the spin belongs to, consistent with **(a)**. Additionally, $\phi$ indicates the angle between the spins on the same zigzag, and, at $J_2 = 0$ it is $\cos^{-1}(1/4)$. As for the zigzag state, the color indicates the sublattice for the collinear state. In the valence-bond state (VBS) the blue bonds indicate where singlets are forming.

phase, there are two possible arrangements of the spins which transform to each other if $\boldsymbol{S}_3$ and $\boldsymbol{S}_4$ are interchanged. Classically, the two different arrangements have chirality of different signs. Due to the lattice geometries used in iDMRG and ED breaking the $C_3$ symmetry of the triangular lattice, the weight of $S(q)$ is not necessarily the same at each of the inequivalent $M$ points.

**Collinear**– In this phase, there are peaks of $S(q)$ at the $M$ points in the Brillouin zone like in the tetrahedral phase. This corresponds to the classical two-spin configuration seen in Fig. S3 where the spins are anti-aligned. Unlike the tetrahedral phase, there is no chirality.

**Zigzag (ZZ)**– The zigzag state has spins ordered along one of the three zigzags and alternating zigzags are anti-aligned. There are peaks of $S(q)$ at the $Y$ points in the Brillouin zone at $(0, \pm\pi)$ as seen in Fig. S6. These points lie

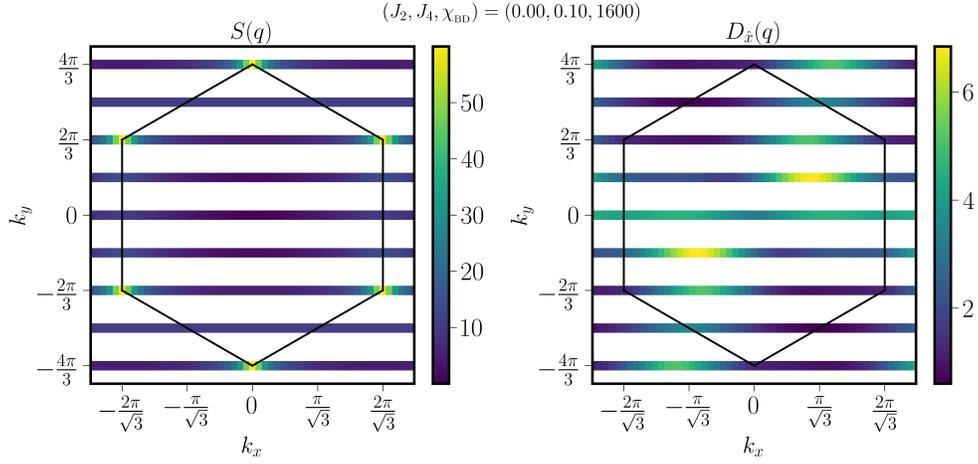

Figure S4. Correlation functions for the 120° order phase from iDMRG. The peaks of $S(q)$ are at the $K$ points in the first Brillouin zone.

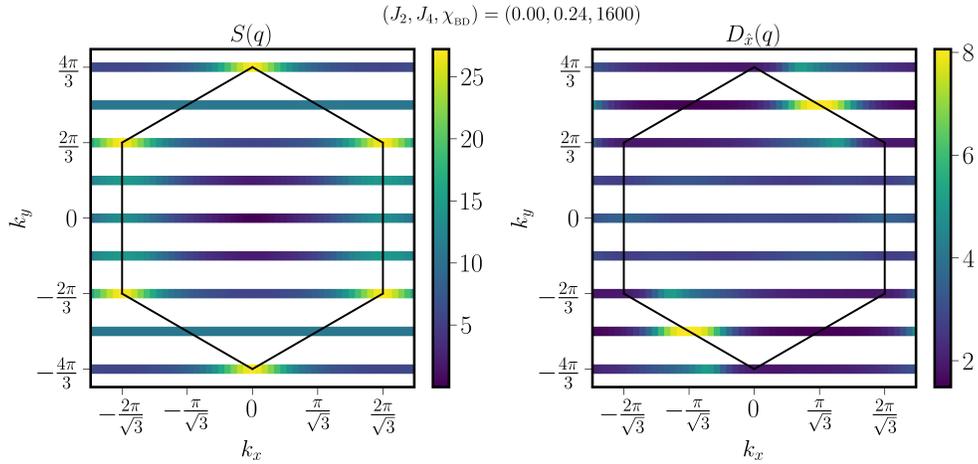

Figure S5. Correlation functions for the valence-bond solid (VBS) state from iDMRG. The peaks of $D_{\bar{x}}(q)$ are at the $M$ points in the first Brillouin zone.

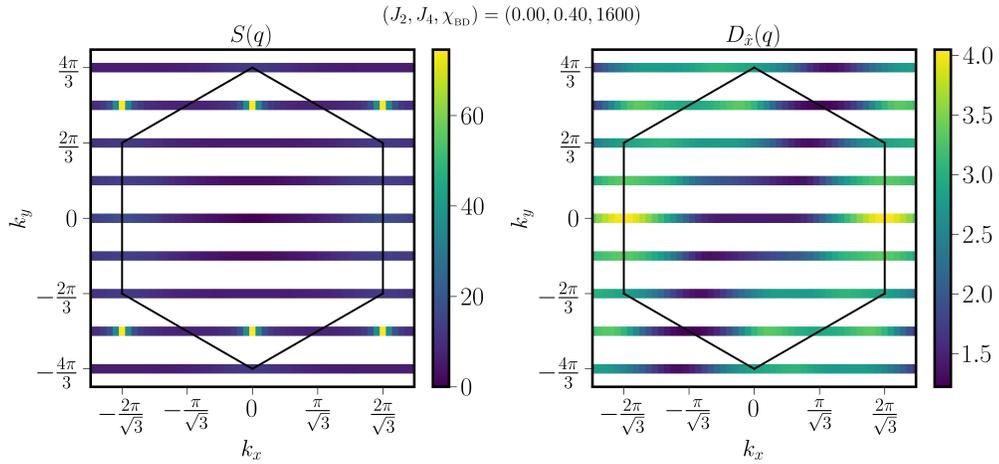

Figure S6. Correlation functions for the zigzag phase from iDMRG. The peaks of $S(q)$ are at the $Y$ points in the first Brillouin zone.

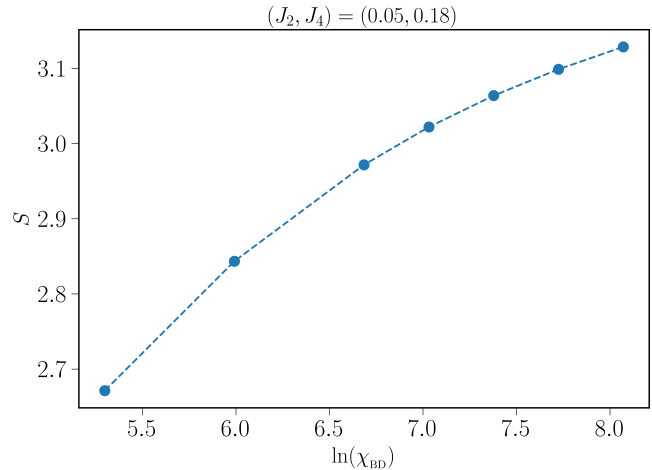

Figure S7. The entropy $S$ vs. the log of the bond dimension $\ln(\chi_{\mathrm{BD}})$ for $(J_2, J_4) = (0.05, 0.18)$ and $L_y = 6$ in the CSL phase. If the spin-liquid were gapless, these should be linearly proportional, but we see deviation from a line indicating the CSL is gapped.

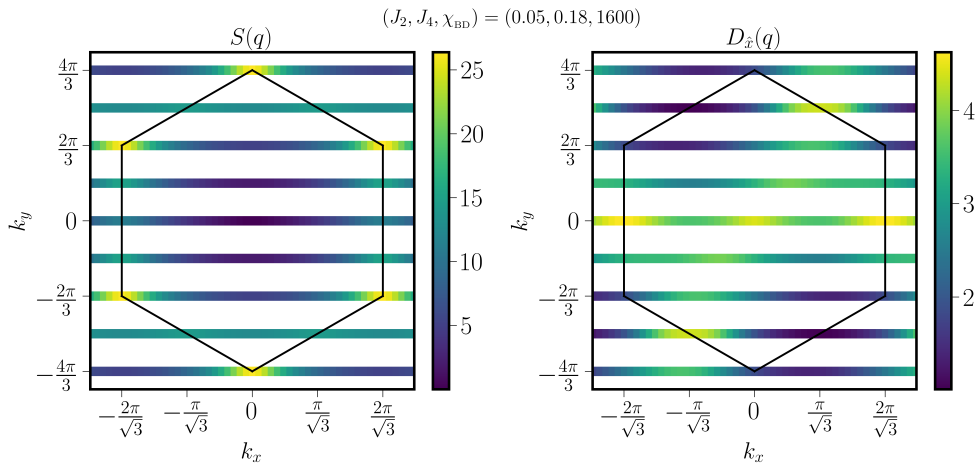

Figure S8. Correlation functions for the CSL phase from iDMRG. Although the correlations still have structure, they are much less strong than in the ordered phases. It is worth noting that the dimer ordering does have features at the $M$ points perhaps indicating that the VBS and CSL state are closely related. However, the VBS state does not break chiral symmetry nor does it have the correct entanglement spectrum counting.

on the middle of the edge of the first Brillouin zone perpendicular to the $k_y$ direction if the real-space unit-cell has a basis of two spins with primitive basis vectors $\hat{y}$ and $2\hat{x} - \hat{y}$ with $\hat{y}$ and $\hat{x}$ are defined as in Fig. 1 in the main text. Within the classical minimization, the particular order is shown in Fig. S3. The angle, $\phi$ between the spins on the same zigzag depends on $J_2$; at $J_2 = 0$, the angle is $\cos^{-1}(1/4)$.

**Valence-bond solid (VBS)** [9]– Within this phase, singlets form along a particular nearest-neighbor bond, and that bonding occurs along a particular axis. An example of the real-space configuration is seen in Fig. S3. This pairing leads to a peak in $D_{\hat{\alpha}}(q)$ at one of the $M$ points where $\hat{\alpha}$ indicates the nearest-neighbor bond that is strongest and the $M$ point indicates how that strong bond is translated around the system. Similar to the other phases, a particular $M$ and $\alpha$ may be selected due to the choice of unit cell. We see this in Fig. S5 where only one of the $M$ points for $D_{\hat{x}}(q)$ is large.

**Spinon Fermi surface (SFS)**– We identify this phase similarly as the phase seen in Ref. [9]. This phase is supposed to have a central charge of $c = 9$. We attempt to extract the central charge, $c$, using the scaling between the entropy $S$ and the correlation length $\xi$ or bond dimension $\chi_{\mathrm{BD}}$ (not shown), but we obtain inconsistent results

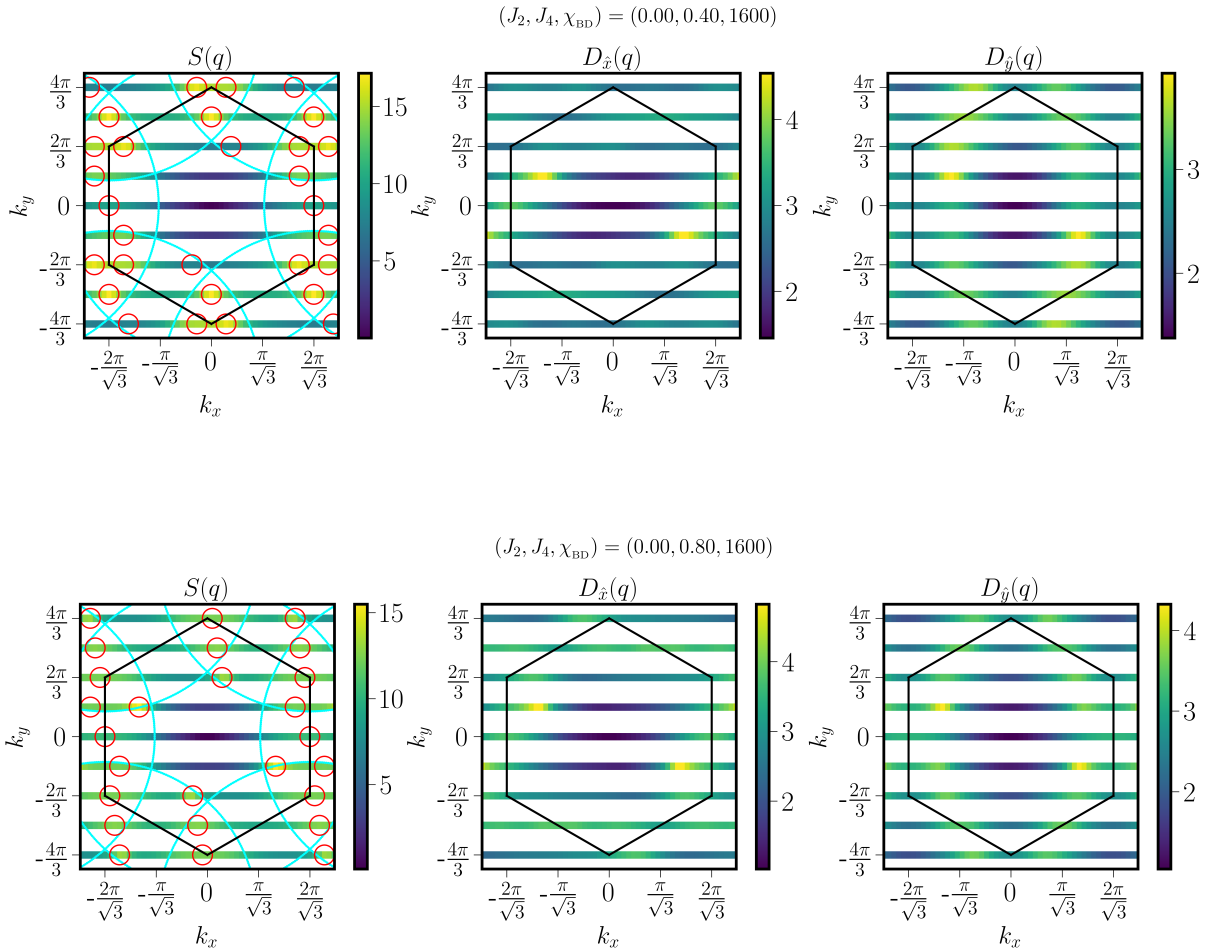

Figure S9. Correlation functions for the putative SFS phase found in Ref. [9] shown at two different parameter points. The cyan lines indicate the $2k_F$ circles approximating the location of scattering from one point to its antipode on the Fermi surface. The red circles indicate the location of maxima in the $k_x$ direction.

depending on whether we use $S \sim (c/6) \ln(\xi)$ or $S \sim 1/(\sqrt{12/c} + 1) \ln(\chi_{\mathrm{BD}})$ [6, 9, 10]. Additionally, the peaks in $S(q)$ are supposed to be near $2k_F$. At two different parameter points, we plot $S(q)$, $D_{\tilde{x}}(q)$, and $D_{\tilde{y}}(q)$. For the $S(q)$ plot, we plot the $2k_F$ surfaces in cyan and identify maxima in the $k_x$ direction with red circles. The Fermi surface is well approximated by a circle, and therefore has a radius $k_F = \sqrt{4\pi/\sqrt{3}} \approx 2.68$ such that its area is half of that of the first Brillouin zone. Although Ref. [9] plots the maxima of $S(q)$, it is not clear how to unambiguously identify maxima (as opposed to saddle points) when $S(q)$ is only defined on the slices shown. Many of the maxima in the $k_x$ direction that we find seem to be in similar locations to Ref. [9], so we expect that we are finding the same state. As the authors of [9] state, however, it is not clear if this state is SFS state because a quantized fractional central charge cannot be extracted with sufficient precision. As we see below, we do not find this state as the ground state within iDMRG for $L_y = 6$ within the range of parameters we focus on, although it appears to be the ground state at larger $J_4$.

**Chiral spin liquid (CSL)**– As mentioned in the main text, the CSL is identified by a non-zero value of the chiral order parameter $\chi$ as well as the proper counting in the entanglement spectrum and a quantized fractional spin-Hall effect. The spin-liquid is gapped (or the quantization would not occur), which we confirm in Fig. S7 by noting that $S$ does not scale linearly with $\ln(\chi_{\mathrm{BD}})$ but instead appears to be leveling off. the As seen in Fig. S8, the peaks of $S(q)$ and $D_{\tilde{x}}(q)$ are significantly smaller than those in the ordered phases that we show, indicating the absence of long range order.

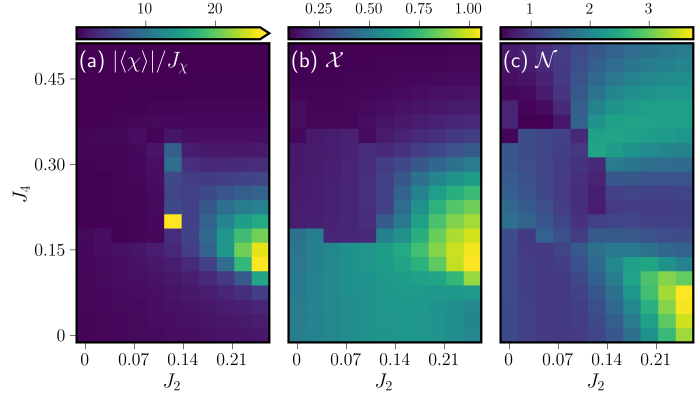

Figure S10. **(a)** We plot $|\langle\chi\rangle|/J_\chi$ vs. $J_2$ and $J_4$ with $J_\chi = 10^{-5}$. There is a single point, $(J_2, J_4) = (0.125, 0.2)$ where $|\langle\chi\rangle|/J_\chi \sim 40$ is much larger than all the other points. We attribute this to the point falling very close to the phase boundary between tetrahedral and VBS states and therefore having a large susceptibility. Within the rest of the phase diagram, however, the largest values occur exactly where $S$(M') is largest. Because the tetrahedral state inherently has a larger susceptibility to chirality [12, 13], this analysis indicates that the $S$(M') peak corresponds to the tetrahedral order. **(b)** We plot the chiral-chiral correlations defined in Eq. (6) which show a significantly larger value in the tetrahedral state than in the collinear state. **(c)** We plot the nematic order parameter Eq. (7), which is larger in the collinear state than the tetrahedral state [12]. We see that it has its peak in the location where $S$(M) is largest.

## EXACT DIAGONALIZATION (ED)

We perform the exact diagonalization in the zero magnetization sector by incorporating translation symmetry and spin-flip symmetry [11]. The periodic boundary conditions can be specified by two vectors indicating how to translate the unit cell to fill the plane. In order for the various orders to be commensurate with our unit cell, we chose the vector parallel to $\hat{y}$ and the vector perpendicular to $\hat{y}$ (see Fig. 1 in the main text). If we instead chose the two vectors parallel to the primitive vectors of the triangular lattice, the 120° order is commensurate only if $L_x$ and $L_y$ are multiples of 3, and ZZ order is commensurate only if $L_x$ and $L_y$ are even and one of them is divisible by 4. Therefore, we cannot choose those periodic boundary conditions with the $(L_y, L_x) = (6, 4)$ unit cell.

**Distinguishing tetrahedral and collinear state in ED-** As mentioned in the main text, we distinguish the tetrahedral and collinear state in three ways. The first is by introducing a small $H_\chi = J_\chi \sum_{\triangle, \triangledown} \boldsymbol{S}_i \cdot (\boldsymbol{S}_j \times \boldsymbol{S}_k)$ term to the Hamiltonian and computing the value of $\langle\chi\rangle$ in the ground state where $\chi = \boldsymbol{S}_i \cdot (\boldsymbol{S}_j \times \boldsymbol{S}_k)$ with $i, j, k$ going clockwise around a triangle and and averaging the expectation over all the triangles. When $J_\chi = 0$, the Hamiltonian does not break TRS and the ED system is finite so $\langle\chi\rangle = 0$. Therefore we compute $|\langle\chi\rangle|/J_\chi$ when $J_\chi = 10^{-5}$ since we expect $|\chi| \sim J_\chi$. The results of this computation are seen in Fig. S10(a). Besides a single point at $(J_2, J_4) = (0.125, 0.2)$ which we explain with a larger susceptibility due to the proximity to the VBS-tetrahedral phase transition, the largest values of $|\langle\chi\rangle|/J_\chi$ is seen to overlap exactly at the point that $S$(M') is large, as seen in Fig. 2(a) in the main text.

The second method is through the computation of the chiral-chiral correlations defined by

$$\mathcal{X} = \sum_{\triangle, \triangledown} \langle [\boldsymbol{S}_0 \cdot (\boldsymbol{S}_{0+\hat{y}} \times \boldsymbol{S}_{0+\hat{x}})][\boldsymbol{S}_i \cdot (\boldsymbol{S}_j \times \boldsymbol{S}_k)] \rangle \tag{6}$$

where $\boldsymbol{S}_0$ is the spin at the bottom left of the unit cell. There are correlations of some strength over the entire bottom half of the phase diagram, including in the 120°, spin liquid and collinear phases. However, the strongest correlations coincide with the points where $S$(M') is largest identifying this as the tetrahedral phase and clearly distinguish it from the collinearly ordered region.

The third way we distinguish the two phases is using a nematic order parameter [12].

$$\mathcal{N} = \sum_{\boldsymbol{x}_i - \boldsymbol{x}_j || \hat{y}} \langle (\boldsymbol{S}_0 \cdot \boldsymbol{S}_{0+\hat{y}})(\boldsymbol{S}_{\boldsymbol{x}_i} \cdot \boldsymbol{S}_{\boldsymbol{x}_j}) \rangle. \tag{7}$$

The sum is taken over all pairs of spins such that $\boldsymbol{x}_i - \boldsymbol{x}_j$ is parallel to $\hat{y}$. The result is shown in Fig. S10(b), and consistently, its value is larger in the collinear phase.

**Overlap with the Kalmeyer-Laughlin state–**

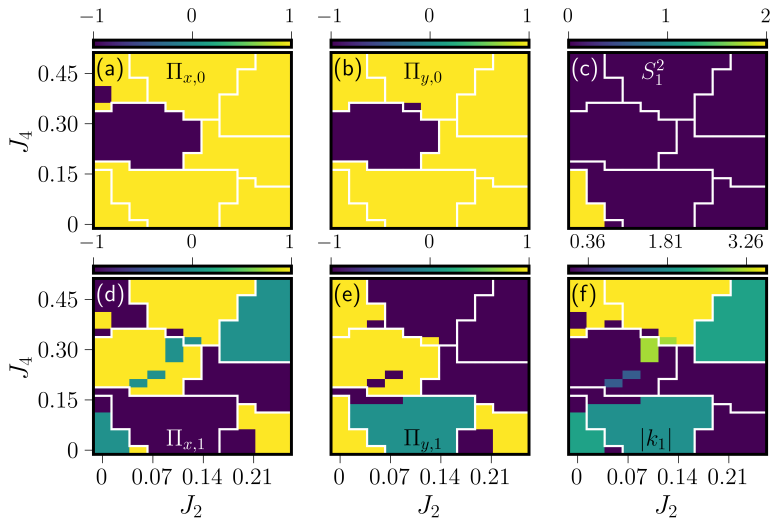

Figure S11. We plot the quantum numbers of the ground state and first excited state. On all plots, we sketch the phases we find in white. Although the phases are not precisely determined by the data (there are a few ambiguous points) they capture the data well. The boundaries we find here are consistent with the order parameters seen in the main text. **(a)**[**(d)**] We plot the quantum number of inversion about the axis perpendicular to the $y$-axis, $\Pi_x$ for the ground state [excited state]. A value of 0 indicates the state is not an eigenstate of $\Pi_x$. **(b)**[**(e)**] We plot the same as for **(a)** and **(d)** buf for inversion about the $y$-axis, $\Pi_y$. **(c)** We plot the total spin squared $S^2$ for the excited state. It is always 0 for the ground state. **(f)** We plot the magnitude of $\vec{k}$ for the excited state. If $\Pi_x$ or $\Pi_y$ fixes $\vec{k}$, then the state is an eigenstate of that operator.

The Kalmeyer-Laughlin state on the torus is given explicitly in the $S_i^z$ basis by [14]

$$\langle S_1^z S_2^z \cdots S_N^z | \psi_k^{\text{KL}} \rangle \sim \delta_s \chi_s \theta \begin{bmatrix} k \\ 0 \end{bmatrix} \left( \sum_{i=1}^N \zeta_i s_i^z, 2\tau \right) \prod_{i<j} \left( \theta \begin{bmatrix} 1/2 \\ 1/2 \end{bmatrix} (\zeta_i - \zeta_j), \tau \right)^{(s_i^z s_j^z + 1)/2} \tag{8}$$

where $s_i^z = \pm 1 = 2S_i^z$, $\delta_s = 1$ if $\sum_{i=1}^N s_i^z = 0$ and is 0 otherwise, $\chi_s = \prod_{j=1}^N (-1)^{(j-1)(s_j^z+1)/2}$, $\zeta_i = (x_i + iy_i)/\omega_1$ for $(x_i, y_i)$ the coordinates of the lattice point with spin $\boldsymbol{S}_i$, $\tau = i\omega_2/\omega_1$, $\omega_1 = N_x\sqrt{3}/2$ is the length of the torus in the $x$ direction, $\omega_2 = N_y$ is the length of the torus in the $y$ direction and

$$\theta \begin{bmatrix} a \\ b \end{bmatrix} (\zeta, \tau) = \sum_{n \in \mathbb{Z}} e^{i\pi\tau(n+a)^2 + 2\pi i(n+a)(\zeta+b)} \tag{9}$$

is the Riemann-theta function. The trivial and semion sector are distinguished by a value of $k = 0$ and $k = 1/2$, respectively. The time-reversal partners to this state occur by changing $\zeta_i \to \zeta_i^*$ (or just complex conjugating the wave function).

We can therefore explicitly compute the wave function, determine which symmetry sector it belongs to (either numerically or by noting it is a spin-0 state and that permuting the order of the spins leads to a factor of the sign of the permutation [14]), and then write down the wave function in that symmetry sector.

**Choosing cut-offs for the phase diagram**– For an ordered state, the excitation spectrum is expected to take a particular form known as the Anderson tower of states [12, 15]. To determine the phase boundaries, then, we use the symmetry sector of the ground state and first excited state. We already incorporated translation symmetry, with quantum number $\vec{k}$ and spin-flip symmetry into the diagonalization. We additionally have the point-group symmetries, which due to our rectangular unit cell is reduced from the usual $D_6$ of the triangular lattice to the group generated by inversion about the $y$ axis, $\Pi_y$ and inversion about the axis perpendicular to $y$, $\Pi_x$.

Since these operations do not, in general, commute with the translation operators, the eigenstates found in ED are only an eigenstate of these operators when the operation fixes $\vec{k}$ (i.e. they belong to the little group of $\vec{k}$) [12, 15]. The ground state is almost always in the $\Gamma$ sector, and, when it is not, it is in the $M$ sector, and therefore the ground state is always an eigenstate of both $\Pi_x$ and $\Pi_y$. We plot their values in Fig. S11(a)–(b).

The first excited state has many different values of $\vec{k}$, as we shown in Fig. S11(c)). If the excited state is degenerate, we plot the value of the state with the smaller value of $|\vec{k}|$. Additionally, the first excited state is often an eigenstate

of $\Pi_x$ and $\Pi_y$, and we plot their eigenvalues in Fig. S11(d)–(e). If $\Pi_x$ or $\Pi_y$ does not fix $\vec{k}$, we use a default value of 0. Finally, we also plot the total spin quantum number of the first excited state $S^2$ in Fig. S11(c).

Although there are some outlying points close to phase boundaries, all of these quantum numbers give a consistent picture of the phase diagram, which we map out over the plots. These are the boundaries we include in Fig. 1(c) in the main text, and the phase boundaries are consistent with the order parameters we measure for each of the phases.

It is worth noting the differences between our phase diagram from ED (Fig. 1 in the main text), and the results of Ref. [2]. We find ordered states like the VBS, the tetrahedral state, and some other dimer ordering at intermediate to large $J_4$ that were not considered within the variational study.

## IDMRG

In this section, we provide details of the iDMRG calculations.

**Setup**– As mentioned in the main text, we use the `TeNPy` library [16] to perform the iDMRG. At a point $(J_2, J_4)$, we create an initial state by starting with a random product state and then perform two sweeps with an added small time-reversal symmetry (TRS) breaking term and a maximum bond dimension of 20. We then set the maximum bond dimension to $\chi_{\rm BD}$, perform 40 sweeps with density matrix mixing [16, 17], and then continue without mixing until the energy difference of subsequent sweeps is $< 10^{-8}$. In general, we can reach a maximum bond dimension of $\chi_{\rm BD} = 3200$ due to the relatively large size of the matrix product operator (MPO) of the Hamiltonian.

By starting with a TRS broken initial state, we can probe whether the resulting ground state spontaneously breaks TRS by checking if the chiral order parameter, $\chi = \langle \boldsymbol{S}_i \cdot (\boldsymbol{S}_j \times \boldsymbol{S}_k) \rangle$, is non-zero at the end of the sweeps. If the value of $\chi$ has decreased to near-zero, the state does not break TRS. However, if the value of $\chi$ increases, the state breaks TRS.

Although `TeNPy` can generate the MPO for Eq. (2) in the main text automatically, the four-spin term is not handled optimally and the resulting bond dimension of the MPO is larger than necessary. We construct the MPO manually using the mapping to a finite state machine and reusing intermediate states in the optimal way [18, 19]. Doing so results in a reduction in MPO size of by a bit more than a factor of 2, which reduces memory cost accordingly.

At large $J_4$, iDMRG has difficulties converging to the correct ground state. To aid in the convergence, after the algorithm converges to a certain state for all $(J_2, J_4)$ points we are considering, we identify the states which seem to have converged to lower energies when compared to surrounding points. For each of those states, we then bias the algorithm by using those states as the initial state and updating the environment prior to optimizing. We then take the lowest energy result from all the biased and unbiased runs as the ground state.

As mentioned above, we compute $S(q)$ and $D_{\rm G}(q)$ as defined in the main text. When we calculate these, we generally compute correlations out to 40 rings along the cylinder.

**Extrapolation**– We follow Refs. [20, 21] and assume that the energy scales as $E_{\chi_{\rm BD}} = E_\infty + A p(\chi_{\rm BD})$ where $p(\chi_{\rm BD})$ $(E_{\chi_{\rm BD}})$ is the truncation error (energy) at bond dimension $\chi_{\rm BD}$, and $A$ is a fitting parameter. In [6], the extrapolation of the chiral order parameter is done by assuming the phenomenological form

$$\chi_{\rm BD} = \chi_\infty + B p(\chi_{\rm BD})^C, \tag{10}$$

for fitting constants $\chi_\infty$, $B$, and $C$. When we use this fitting form, we find that at $J_2 = 0.0$ and $J_4 = 0.19$, this extrapolation leads to a *negative* value of $|\chi|$. Therefore, we instead use the alternative form

$$\chi_{\chi_{\rm BD}} = \chi_\infty + \frac{B}{\chi_{\rm BD}} + \frac{C}{\chi_{\rm BD}^2}, \tag{11}$$

which is just a Taylor expansion in the variable $1/\chi_{\rm BD}$ around $\chi_{\rm BD} \to \infty$. We use this formula to find the infinite bond dimension value of $\chi$ as plotted in the main text, and we use the energy extrapolation below.

We compare the two extrapolation methods for $\chi_\infty$ in Fig. S12. As can be seen, the two possible fitting functions do not disagree too much deep in the CSL phase, but there is some disagreement at the boundaries. It is important to note that the only clearly nonsensical results of Eq. (11) occur at points where the nature of the ground state changes as a function of bond dimension.

**ZZ vs. SFS**– As noted in the main text, we find that the ZZ phase seems to be the ground state in ED and iDMRG with $L_y = 6$. To verify that this holds for all bond dimensions, we extrapolate the energy to infinite bond dimension. We show the results in Fig. S13. As we see, the ZZ appears to extrapolate to lower energy at $(J_2, J_4) = (0.0, 0.4)$ and the SFS is probably lower energy at $(J_2, J_4) = (0.0, 0.8)$.

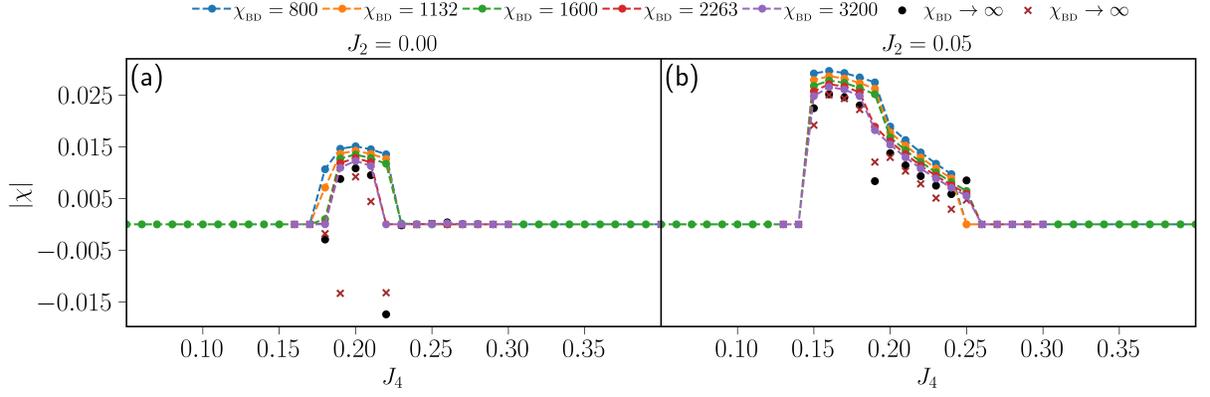

Figure S12. We plot $|\chi|$ vs. $J_4$ for two values of $J_2$, as indicated by the title. We show the results of the two possible extrapolation equations. The brown crosses result from using Eq. (10) while using Eq. (11) leads to the black circles. Note that the two produce close results in most cases, but Eq. (10) generates a negative $|\chi|$ at $J_2 = 0.05$ and $J_4 = 0.19$ for no clear reason. Eq. (11) only leads to $|\chi| < 0$ when the nature of the ground state seems to have changed at intermediate bond dimension.

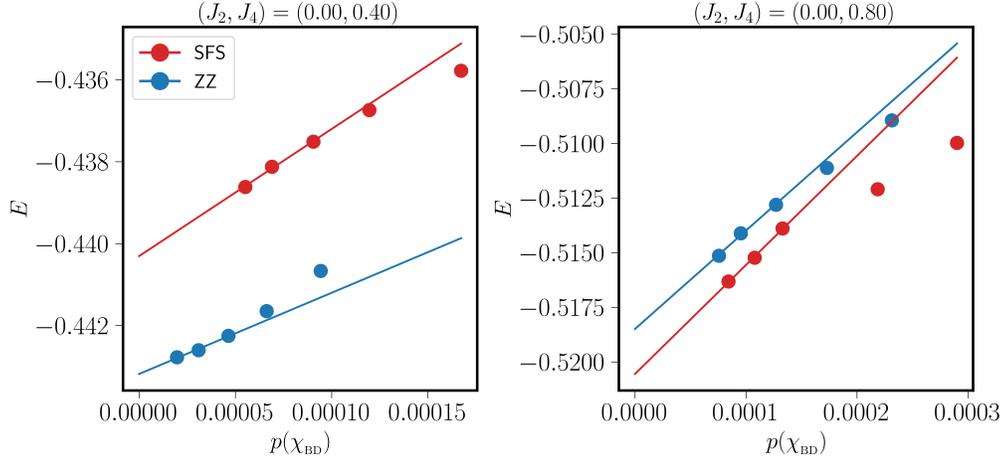

Figure S13. We plot the energy vs. truncation error at different bond dimensions for the zigzag (ZZ) and putative spinon Fermi surface (SFS) states whose correlators are seen in Fig S6 and Fig. S9. We attempt to extrapolate to infinite bond dimension by fitting the energy to the form $E_{\chi_{\mathrm{BD}}} = E_\infty + Ap(\chi_{\mathrm{BD}})$ for some fitting parameters $E_\infty$ and $A$. We only use the largest three bond dimensions for this fitting. In both plots $J_2 = 0$. For the $J_4 = 0.4$ plot, it appears the ZZ state is lower in energy, but for $J_4 = 0.8$ it seems like the SFS is lower energy. The bond dimensions used are 800, 1132, 1600, 2263, and 3200 and $L_y = 6$.

**The $J_4 = 0.16$ slice**– In the main text, we only showed data from performing iDMRG at fixed $J_2$. If we instead consider a fixed $J_4$, we find that we need to increase the number of rings in the unit cell to $L_x = 4$ instead of $L_x = 2$ used in the main text in order for the tetrahedral order to be commensurate with the unit cell. The increased unit cell size makes it more difficult to go to larger bond dimension.

Nevertheless, we include data where we vary $J_2$ and keep $J_4 = 0.16$ for bond dimension $\chi_{\mathrm{BD}} \leq 1600$ in Fig. S14. We find $120°$ order for $J_2 \leq 0.025$, the CSL for $0.025 \leq J_2 \lesssim 0.175$ and the tetrahedral order for $J_2 \gtrsim 0.175$, although the phase boundary between the CSL and the tetrahedral order phase is less clear.

**Entanglement spectrum and spin pumping**– See e.g. Ref. [22] (in particular, Appendix C of the arXiv version [23]) or the `TeNPy` documentation for background information. The entanglement spectrum is computed by finding the eigenvalues of the reduced density matrix for half of the cylinder. The momentum $k_\alpha$ of the Schmidt states is determined by constructing a mixed density matrix of the ground state wave-function and the ground state wave-function translated by one unit vector in the $\hat{y}$ direction. The momentum can then be read off from the phase $e^{ik_\alpha}$ of the eigenvalues of the mixed density matrix. There is an undefined phase in this procedure, which corresponds to shifting all $k$ by a constant and we fix this gauge freedom by setting $k = 0$ for the lowest Schmidt state. An alternative approach would be to fix $\sum_\alpha s_\alpha^2 e^{ik_\alpha}$ to be real with $s_\alpha$ being the Schmidt values, which can reveal information about

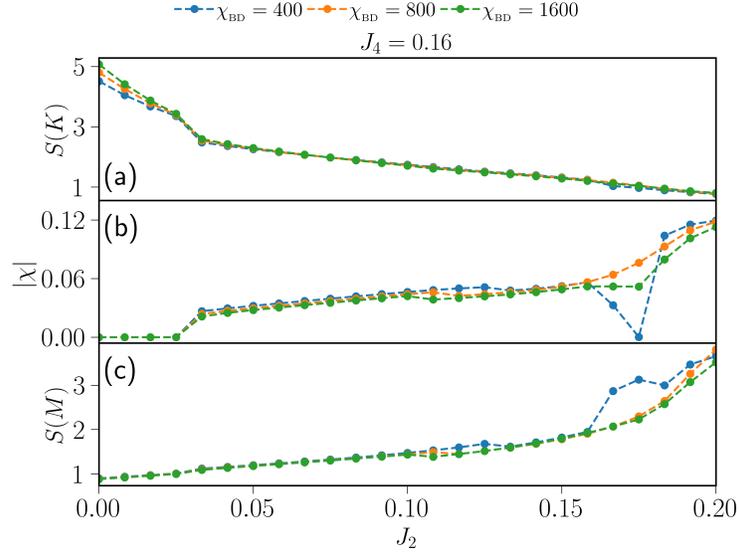

Figure S14. We plot the observables indicating various orders that we extract from ground state wave function from iDMRG for the $L_y = 6$ cylinder with a four ring unit cell, and $J_4 = 0.16$. In **(a)**, we show the spin-spin correlations at the $K$ point. The jump around $J_2 = 0.025$ corresponds to the transition from the $120°$ into the CSL phase. **(b)** We plot the absolute value of $\chi = \langle \boldsymbol{S}_i \cdot (\boldsymbol{S}_j \times \boldsymbol{S}_k) \rangle$ averaged over all triangles of the lattice from the iDMRG results at varying bond dimension, $\chi_{\mathrm{BD}}$. The transition into the CSL is clearly visible as the onset of a finite $\chi$. The transition into the tetrahedral order can be identified by the changing slope of $|\chi|$ around $J_2 \approx 0.175$. As in the main text, the slight jump in the non-zero value of $|\chi|$ at around $J_2 = 0.1$ corresponds to whether the trivial or semion sector of the KL state is the ground state. In **(c)**, we show the spin-spin correlations at the $M$ point. The features are much less clear, but the onset of the tetrahedral order can be seen from the change in slope at larger $J_2$.

the anyons in the system [24–26]. Based on those studies, however, it would likely require $L_y \gtrsim 12$ to reliably extract that information.

The spin being pumped through the cylinder is computed via $\sum_\alpha s_\alpha^2 s_{z,\alpha}$ where $s_\alpha$ are the Schmidt values, $s_{z,\alpha}$ are the spin charges of the Schmidt states on the left half of the system and the sum goes over the $\chi_{\mathrm{BD}}$ eigenvalues of the reduced density matrix. Since we are implementing twisted periodic boundary conditions uniformly over the cylinder, our spin creation/annihilation operators are related to those with the twist happening entirely at the boundary via $S_j^+ \mapsto S_j^+ e^{i y_j \theta / L_y}$, where $y_j$ is the distance around the circumference of the cylinder of the point $j$, and $\theta$ is the twist. When we compute the entanglement spectrum, therefore, the momenta are shifted by $k \to k - s_z \theta / L_y$, so we undo this shift after the $k$ values are computed. Additionally, the charges themselves are shifted by $s_z \to s_z + \sigma_{xy}^{\mathrm{spin}} \theta / (2\pi)$ due to the accumulation of spin by the finite spin Hall conductivity.

In addition to the $L_y = 8$ data included in the main text, we give two more examples of spin pumping. The first in Fig. S15 is the $L_y = 6$ equivalent of Fig. 4 in the main text. We see the same features–the correct level counting for the KL state, the quantized spin-Hall conductivity, and the change in topological sector.

We additionally attempt the same spin-pumping experiment in the tetrahedral state in Fig. S16. There are many clear differences, but the most important is that the state at $\theta = 0$ is the same as the state at $\theta = 2\pi$ (albeit with opposite chirality) indicating no topological degeneracy. Additionally, the spin Hall conductivity is zero instead of quantized to a finite value.

## MEAN FIELD ARGUMENTS

The tendency for spontaneous chiral symmetry breaking for $J_4 > 0$ and the similarity of the overall structure of the phase diagram to the phase diagram of the $J_1 - J_2 - J_\chi$ Hamiltonians studied in [12, 13] suggests that we can rewrite the Hamiltonian in a way that makes the spontaneous TRS breaking manifest. In this section, we provide a mean-field argument that does exactly that. Our starting point is the $J_2 = 0$ Hamiltonian and the observation that, for spin-1/2s,

$$\hat{\chi}_{ijkl}^2 = \mathcal{O}_\triangle(i,j,l) \cdot \mathcal{O}_\triangledown(k,l,j) + \mathcal{O}_\triangledown(k,l,j) \cdot \mathcal{O}_\triangle(i,j,l) = -S_i \cdot S_k + 2(S_i \cdot S_j)(S_k \cdot S_l) + 2(S_i \cdot S_l)(S_k \cdot S_j), \quad (12)$$

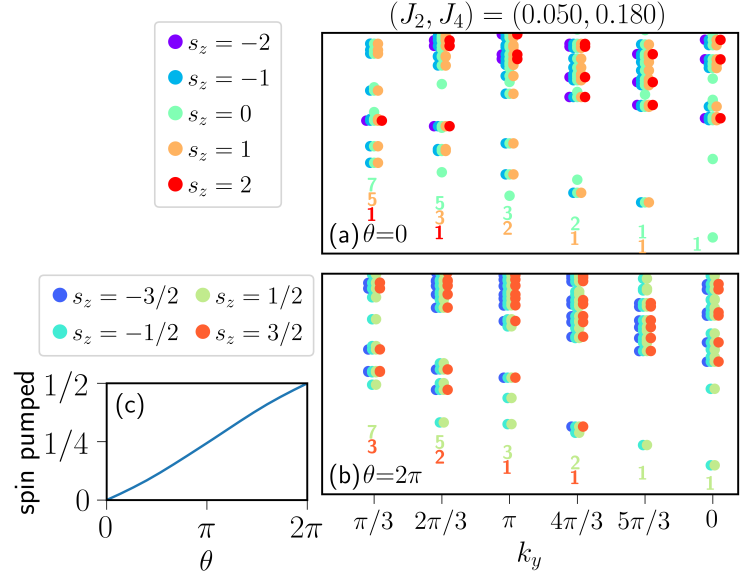

Figure S15. We plot the equivalent of Fig. 4 in the main text but for the $L_y = 6$ cylinder and with bond dimension $\chi_{BD} = 800$. The main difference is that the spin-pumping is a little less linear (though still quantized at $\theta = 2\pi$), and that fewer $k$ values in the entanglement spectrum allow us to resolve fewer terms in the series counting the edge states.

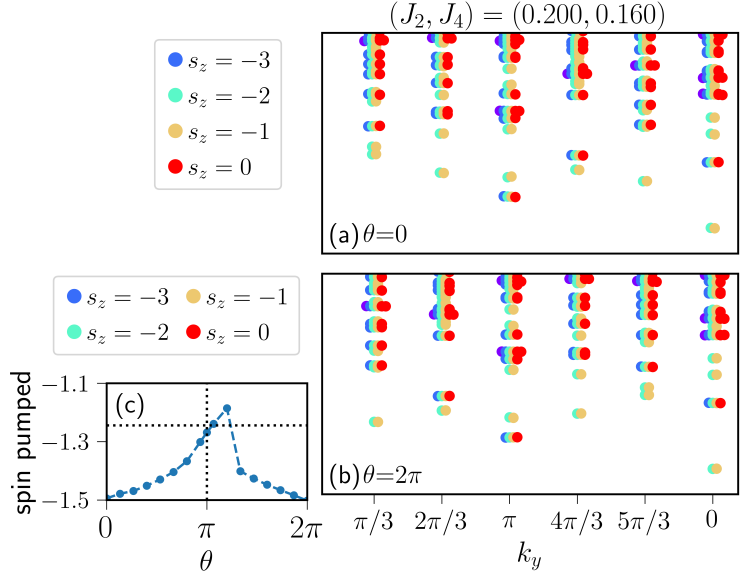

Figure S16. We plot the equivalent of Fig. 4 in the main text but for the $L_y = 6$ cylinder and for a tetrahedral ground state. The entanglement spectrum does not follow the pattern for the KL state, the spin pumping results in no net change of spin accumulation, and the state at $\theta = 0$ and $\theta = 2\pi$ is the same but with opposite chirality. We attribute this switching of chiralities at around $\theta = 1.3\pi$ to an interplay between finite size effects and finite flux boundary conditions that affects the energy of the states. In **(c)** the dotted black lines indicate a value of $\theta = \pi$ and of $1/4$ spin being accumulated–if the state had a quantized Hall conductivity, then the point where the two lines cross should be on the blue curve. For this plot, $\chi_{BD} = 1600$.

with $\mathcal{O}_\triangle(i,j,l) = 2S_i \cdot (S_j \times S_l)$ around the triangle oriented in the same way (e.g. clockwise) and $i$ and $k$ are any two next-nearest neighbors. Additionally note

$$
\begin{aligned}
\hat{\chi}^4_{ijkl} = (\hat{\chi}^2_{ijkl})^2 =& \frac{9}{16} - \frac{1}{4}(\boldsymbol{S}_i \cdot \boldsymbol{S}_k) - \frac{1}{4}(\boldsymbol{S}_j \cdot \boldsymbol{S}_l) - \frac{5}{8}\left[(\boldsymbol{S}_i \cdot \boldsymbol{S}_j) + (\boldsymbol{S}_k \cdot \boldsymbol{S}_l) + (\boldsymbol{S}_i \cdot \boldsymbol{S}_l) + (\boldsymbol{S}_k \cdot \boldsymbol{S}_j)\right] \\
&+ \frac{1}{2}\left[(\boldsymbol{S}_i \cdot \boldsymbol{S}_j)(\boldsymbol{S}_k \cdot \boldsymbol{S}_l) + (\boldsymbol{S}_i \cdot \boldsymbol{S}_l)(\boldsymbol{S}_k \cdot \boldsymbol{S}_j)\right] + 2(\boldsymbol{S}_i \cdot \boldsymbol{S}_k)(\boldsymbol{S}_j \cdot \boldsymbol{S}_l) \\
\hat{\chi}^6_{ijkl} = (\hat{\chi}^2_{ijkl})^3 =& \frac{15}{64} - \frac{11}{16}(\boldsymbol{S}_j \boldsymbol{S}_l) - \frac{15}{16}(\boldsymbol{S}_i \cdot \boldsymbol{S}_k) - \frac{5}{16}\left[(\boldsymbol{S}_i \cdot \boldsymbol{S}_j) + (\boldsymbol{S}_k \cdot \boldsymbol{S}_l) + (\boldsymbol{S}_i \cdot \boldsymbol{S}_l) + (\boldsymbol{S}_k \cdot \boldsymbol{S}_j)\right] \\
&+ \frac{13}{4}\left[(\boldsymbol{S}_i \cdot \boldsymbol{S}_j)(\boldsymbol{S}_k \cdot \boldsymbol{S}_l) + (\boldsymbol{S}_i \cdot \boldsymbol{S}_l)(\boldsymbol{S}_k \cdot \boldsymbol{S}_j)\right] + \frac{5}{4}(\boldsymbol{S}_i \cdot \boldsymbol{S}_k)(\boldsymbol{S}_j \cdot \boldsymbol{S}_l)
\end{aligned}
\tag{13}
$$

We can then rewrite the $J_4$ term without generating an additional $J_2$ term using the linear combination

$$
\begin{aligned}
-\frac{39}{88}\hat{\chi}^2_{ijkl} - \frac{21}{22}\hat{\chi}^4_{ijkl} + \frac{8}{11}\hat{\chi}^6_{ijkl} =& \left[(\boldsymbol{S}_i \cdot \boldsymbol{S}_j)(\boldsymbol{S}_k \cdot \boldsymbol{S}_l) + (\boldsymbol{S}_i \cdot \boldsymbol{S}_l)(\boldsymbol{S}_k \cdot \boldsymbol{S}_j)\right] - (\boldsymbol{S}_i \cdot \boldsymbol{S}_k)(\boldsymbol{S}_j \cdot \boldsymbol{S}_l) \\
&- \frac{129}{352} + \frac{65}{176}\left[(\boldsymbol{S}_i \cdot \boldsymbol{S}_j) + (\boldsymbol{S}_k \cdot \boldsymbol{S}_l) + (\boldsymbol{S}_i \cdot \boldsymbol{S}_l) + (\boldsymbol{S}_k \cdot \boldsymbol{S}_j)\right] - \frac{23}{88}(\boldsymbol{S}_j \cdot \boldsymbol{S}_l).
\end{aligned}
\tag{14}
$$

This rewriting leads to a change in the strength of $J_1$ ($\tilde{J}_1 = J_1 - \frac{107}{88}J_4$), which therefore rescales $J_2/J_1 \to J_2/(J_1 - \frac{107}{88}J_4)$ and similar for $J_4$. Otherwise, we end up with the Hamiltonian

$$
\begin{aligned}
H =& J_1 \sum_{\langle ij \rangle} S_i \cdot S_j + J_2 \sum_{\langle\langle ij \rangle\rangle} S_i \cdot S_j + J_4 \sum_{\langle i,j,k,l \rangle} (S_i \cdot S_j)(S_k \cdot S_l) + (S_i \cdot S_l)(S_k \cdot S_j) - (S_i \cdot S_k)(S_j \cdot S_l) \\
=& \tilde{J}_1 \sum_{\langle ij \rangle} S_i \cdot S_j + J_2 \sum_{\langle\langle ij \rangle\rangle} S_i \cdot S_j + J_4 \sum_{\langle i,j,k,l \rangle} \left[ -\frac{39}{88}\hat{\chi}^2_{ijkl} - \frac{21}{22}\hat{\chi}^4_{ijkl} + \frac{8}{11}\hat{\chi}^6_{ijkl} \right] + 3NJ_4\frac{129}{352}
\end{aligned}
\tag{15}
$$

where $N$ is the number of sites.

Then, because the system spontaneously generates an expectation value of $\chi = \langle O_\triangle(i,j,l) \rangle/2 = \langle O_\triangledown(k,l,j) \rangle/2 = \langle O_\triangle(i,j,l) + O_\triangledown(k,l,j) \rangle$, we mean-field decouple the $\hat{\chi}^2_{ijkl}$ terms. This is done by rewriting

$$
\hat{\chi}^2_{ijkl} = 8\chi(\chi + \epsilon_\triangle + \epsilon_\triangledown) + \epsilon_\triangle\epsilon_\triangledown + \epsilon_\triangledown\epsilon_\triangle,
\tag{16}
$$

with $\epsilon_\triangle = \mathcal{O}_\triangle(i,j,l)/2 - \chi$, expanding $(\tilde{\chi}^2_{ijkl})^n$, and keeping only to order $\epsilon$.

In the end, we arrive at the Hamiltonian

$$
\begin{aligned}
H =& \tilde{J}_1 \sum_{\langle ij \rangle} \boldsymbol{S}_i \cdot \boldsymbol{S}_j + J_2 \sum_{\langle\langle ij \rangle\rangle} \boldsymbol{S}_i \cdot \boldsymbol{S}_j + \underbrace{3J_4\left[-\frac{39}{11}\chi - \frac{21}{11}8^2\chi^3 + \frac{3}{11}8^4\chi^5\right]}_{J'_\chi} \sum_{\triangle,\triangledown} \boldsymbol{S}_i \cdot (\boldsymbol{S}_j \times \boldsymbol{S}_k) \\
&+ 3NJ_4\frac{129}{352} - 3NJ_4\left[-\frac{39}{11}\chi^2 - \frac{63}{22}8^2\chi^4 + \frac{5}{11}8^4\chi^6\right]
\end{aligned}
\tag{17}
$$

To determine whether there are self-consistent solutions, we numerically study the Hamiltonian

$$
H = \sum_{\langle ij \rangle} S_i \cdot S_j + J_2 \sum_{\langle\langle ij \rangle\rangle} S_i \cdot S_j + J_\chi \sum_{\triangle,\triangledown} \boldsymbol{S}_i \cdot (\boldsymbol{S}_j \cdot \boldsymbol{S}_k)
\tag{18}
$$

to determine the function $\chi(J_\chi)$. For simplicity, we choose to set $J_2 = 0.05$. We show in Fig. S17 that for $J_4 \sim 0.13$, there are self-consistent solutions where the value of $\chi(J_\chi)$ reproduces the value of $J'_\chi = J_\chi$ in Eq. (17) and the solution has lower energy than the trivial $\chi = 0$ solution (note, however, that $J_2$ in Eq. (17) has a value $J_2/[1 - (107/88)J_4] = 0.05$). Additionally, at this value of $J_\chi$ the ground state is the CSL. It is worth noting that $\chi \sim 0.1$ is much larger than we find numerically, consistent with the expectation that fluctuations will decrease the actual order parameter from its mean-field value, and the range of values of $J_4$ that produce consistent solutions is different than what we find numerically, but those values are not far from what we see numerically for the original Hamiltonian in the first line of Eq. (15).

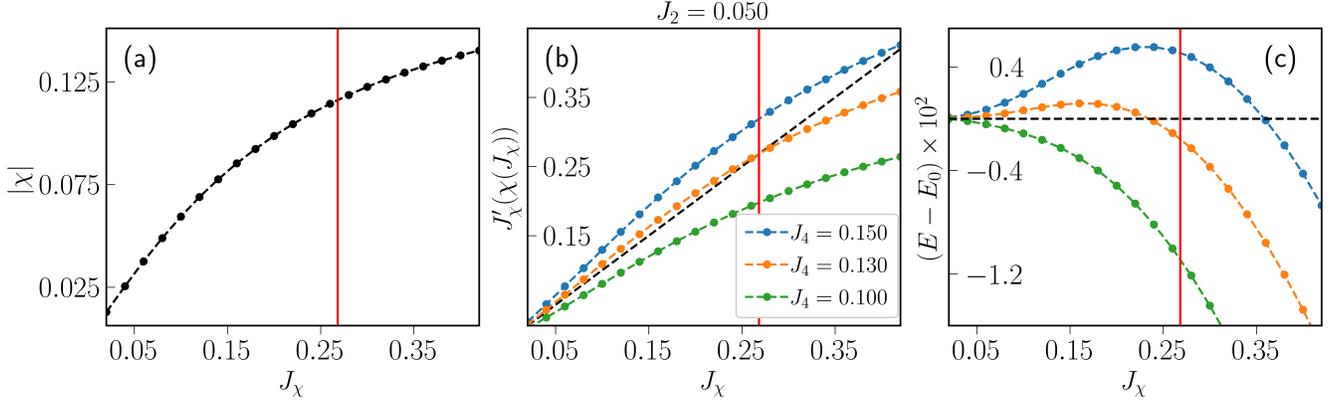

Figure S17. iDMRG data showing the self-consistent solution from the mean-field decoupled Hamiltonian (17). In **(a)**, we plot $\chi = \langle \boldsymbol{S}_i \cdot (\boldsymbol{S}_j \times \boldsymbol{S}_k) \rangle$ averaged over all triangles vs. $J_\chi$ at $\chi_{\text{BD}} = 1600$ (the value of $\chi$ seems to already be converged). In **(b)**, we insert the $\chi(J_\chi)$ function from (a) into Eq. (17) with $\tilde{J}_1 = 1$ to find the resulting value of $J'_\chi$. The different colors indicate different values of $J_4$, and the dashed black line follows $J'_\chi = J_\chi$, which would indicate a consistent solution. There is a solution with $J_4 = 0.13$ with $J_\chi \approx 0.268$. Since $J_\chi \gtrsim 0.2$, the ground state is the KL state [12, 13]. In **(c)**, we plot the energy of ground state using the energy resulting from iDMRG on Eq. (18) plus the shift due to the non-zero value of $\chi$ in Eq. (17). The black dashed line indicates the ground state energy when $J_\chi = 0$. At the self-consistent point with $\chi \neq 0$, the energy is lower.

Finally, we comment on the "uniqueness" of this procedure. The two main choices we had to make are the definition of $\hat{\chi}^2_{ijkl}$ and the rewriting of $J_4$ without generating a next-nearest neighbor term. Since $\hat{\chi}^2$ must include four sites, the main alternative choice is $\bar{\chi}^2_{ijkl} = [\mathcal{O}_\triangle(i,j,k) + \mathcal{O}_\triangledown(j,l,k)]^2$. However, this choice does not work since $\bar{\chi}^2_{ijkl}$ is not linearly independent of $\bar{\chi}^4_{ijkl}$ and $\bar{\chi}^2_{ijkl}$. On the other hand, if we were to not include the $\hat{\chi}^6_{ijkl}$ term (or the $\bar{\chi}^6_{ijkl}$ term), we can replace the $J_4$ term while generating an additional $J_2$, but this produces no self-consistent solution. Finally, one can also find a combination of $\hat{\chi}^2_{ijkl}$, $\hat{\chi}^4_{ijkl}$ and $\hat{\chi}^6_{ijkl}$ that rewrites the $J_2$ term without generating a $J_4$ term. Applying the same analysis as to the above rewriting, we still find self-consistent solutions and we may expect that this would mean the CSL is the ground state for some parameters of the $J_1 - J_2$ Hamiltonian. This conclusion is the same as Ref. [27], however, when studied with numerical methods, it seems that spontaneous TRS breaking does not occur when $J_4 = 0$.

---



\end{document}